\documentstyle[12pt,epsf,epsfig]{article}
\newcommand{\eg}{{\it e.g. }}
\newcommand{\mrm}[1]{\mbox{\rm #1}}
\newcommand{\beq}{\begin{equation}}
\newcommand{\eeq}{\end{equation}}
\newcommand{\bea}{\begin{eqnarray}}
\newcommand{\eea}{\end{eqnarray}}
\newcommand{\rfn}[1]{(\ref{#1})}

\newcommand{\Eq}[1]{Eq.(\ref{#1})}
\newcommand{\nn}{\nonumber}

\def\lsim{\mathrel{\vcenter{\hbox{$<$}\nointerlineskip\hbox{$\sim$}}}}

\def\m12{m_{1\!/2}}
\def\tb{\tan\beta}

\begin{document}
\begin{titlepage}
\pagestyle{empty}
\baselineskip=21pt
\rightline{hep-ph/0109125}
\rightline{CERN--TH/2001-236}     
\rightline{KEK-TH-779}
\vskip 0.25in
\begin{center}
{\large{\bf 
CP Violation in the Minimal Supersymmetric Seesaw Model
}}
\end{center}
\begin{center}
\vskip 0.25in
{{\bf John Ellis}$^1$,
{\bf Junji Hisano}$^{1,2}$,
{\bf Smaragda Lola }$^1$ and
{\bf Martti Raidal}$^{1,3}$
\vskip 0.15in
{\it
$^1${CERN, Geneva 23, CH-1211, Switzerland}\\
$^2${Theory Group, KEK, Oho 1-1, Tsukuba, Ibaraki 305-0801, Japan}\\
$^3${National Institute of Chemical Physics and Biophysics, \\ 
Tallinn 10143, Estonia 
}\\
}}
\vskip 0.25in
{\bf Abstract}
\end{center}
\baselineskip=18pt \noindent

We study CP violation in the lepton sector of the supersymmetric extension
of the Standard Model with three generations of massive singlet neutrinos
with Yukawa couplings $Y_\nu$ to lepton doublets, in a minimal seesaw
model for light neutrino masses and mixing. This model contains six
physical CP-violating parameters, namely the phase $\delta$ observable in
oscillations between light neutrino species, two Majorana phases
$\phi_{1,2}$ that affect $\beta \beta_{0 \nu}$ decays, and three
independent phases appearing in ${Y_\nu}{Y_\nu}^\dagger$, that control the
rate of leptogenesis. Renormalization of the soft supersymmetry-breaking
parameters induces observable CP violation at low energies, including
T-odd asymmetries in polarized $\mu\to eee$ and $\tau \to \ell \ell \ell$
decays, as well as lepton electric dipole moments. In the
leading-logarithmic approximation in which the massive singlet neutrinos
are treated as degenerate, these low-energy observables are sensitive via
${Y_\nu}^\dagger{Y_\nu}$ to just one combination of the leptogenesis and
light-neutrino phases. We present numerical results for the T-odd
asymmetry in polarized $\mu\to eee$ decay, which may be accessible to
experiment, but the lepton electric dipole moments are very small in this
approximation. To the extent that the massive singlet neutrinos are not
degenerate, low-energy observables become sensitive also to two other
combinations of leptogenesis and light-neutrino phases, in this minimal
supersymmetric seesaw model.

\vfill
\leftline{CERN--TH/2001-236}
\leftline{September 2001}
\end{titlepage}
\baselineskip=18pt


\section{Introduction}

The solar \cite{sksol,sno} and atmospheric \cite{skatm} neutrino
anomalies, which imply the existence of non-zero masses for the light
neutrinos, provide the first experimental evidence for the existence of
physics beyond the Standard Model (SM). A minimal extension of the SM
includes three very heavy singlet neutrinos $N_{i}^c$, whose Yukawa
couplings ${Y_\nu}$ to the light neutrinos explain naturally the smallness
of their masses, via the seesaw mechanism \cite{seesaw}. At the same
time, the electroweak scale must be stabilized against large radiative
corrections. In particular, after introducing right-handed neutrinos, a
quadratically-divergent contribution to the Higgs boson mass proportional
to $M_{N}^2$ has to be cancelled. This is most commonly achieved by
supersymmetrizing the theory, leading to the minimal supersymmetric
extension of the Standard Model (MSSM) with singlet neutrinos. 

Neutrino-flavour mixing originates from off-diagonal components in the
Yukawa interaction $N^c{Y_\nu}LH_2$, in a basis where the charged-lepton
and singlet-neutrino mass matrices are real and diagonal. Renormalization
effects due to this interaction also induce flavour mixings in the soft
supersymmetry-breaking slepton mass terms~\cite{bm}. This may lead to
observable rates for charged-lepton flavour-violating (LFV)  processes
such as $\mu\to e\gamma$, $\mu$-$e$ conversion in nuclei, $\mu\to eee$ and
$\tau \to 3 \ell$~\cite{review,h1,ci,nlfv}, where $\ell = e, \mu$ denotes
a generic light charged lepton. LFV is also observable in principle in
rare kaon decays, but at rates that are likely to be far below the current
bounds~\cite{kaons}. 

In general, ${ Y_\nu}$ is complex, leading to CP violation in neutrino
oscillations and in the induced rare LFV processes, as well as in Majorana
phases for the light neutrinos and in electric dipole moments for the
charged leptons. The existence of CP violation in ${ Y_\nu}$ is also
required if the observed baryon asymmetry in the Universe originated in
leptogenesis~\cite{lepto}.  The purpose of this paper is to clarify the
relations between these different manifestations of CP violation in the
lepton sector, and to present numerical estimates of the T-odd
CP-violating asymmetry $A_T$ in $\mu\to eee$ decay, the electric dipole
moments of the electron and muon. We argue that measurements of CP
violation using charged leptons, combined with CP violation in the 
light-neutrino sector, in principle enable the leptogenesis phases to be
extracted - within the framework of the minimal supersymmetric seesaw
model. 

If the solar-neutrino mass-squared difference $\Delta m^2_{sol}$ and the
element $U_{e3}$ of the Maki-Nakagawa-Sakita (MNS) neutrino-mixing matrix $U$
are not too small, the CP-violating phase $\delta$ in $U$, which is
analogous to the Cabbibo-Kobayashi-Maskawa (CKM) phase in the quark
sector, can be measured via CP- and T-violating~\cite{cpt} observables in
neutrino oscillations using a neutrino factory or possibly a low-energy
neutrino superbeam.  The recent SNO result~\cite{sno} encourages this
possibility, since it further favours the large-mixing-angle (LMA) solution to
the solar-neutrino deficit~\cite{snofit}. 

As mentioned above, processes that violate charged-lepton flavour can
provide important complementary information on the leptonic
CP-violating phases. These may be measured using intense sources of
stopped muons. The SINDRUM II experiment is designed to be sensitive
to ${\cal B}(\mu Ti\to e Ti) \sim 10^{-14}$ \cite{sindrum2}, and
the MECO project would be sensitive to ${\cal B}(\mu Al \to e Al)
\sim 10^{-16}$ \cite{meco}. The experiment with the sesitivity ${Br}(\mu\to
e\gamma)\sim 10^{-14}$ is proposed at PSI \cite{PSI}. The PRISM
project~\cite{review,PRISM} and the front ends of neutrino factories
now under consideration at CERN~\cite{nf} and elsewhere will provide
beams of low-energy muons that are more intense by several orders of
magnitude than any of the present facilities. This will enable the
construction of stopped-muon experiments able to probe LFV processes
with sensitivities ${Br}(\mu\to e\gamma) \sim 10^{-15}$, 
${Br}(\mu\to eee) \sim 10^{-16}$. The latter sensitivity opens the way to
measuring the T-odd, CP-violating asymmetry $A_T(\mu\to eee)$.

A measurement of the CP-violating electric dipole moment (EDM) of the muon
with a sensitivity $d_\mu \sim 5\times 10 ^{-26}$ e cm would also be
possible~\cite{nf}.  However, because the Yukawa coupling constants
${Y_\nu}$ appear in the renormalization-group equations (RGEs) only in the
Hermitian combination ${ Y_\nu^\dagger Y_\nu}$, CP-violating phases are
induced only in the off-diagonal terms of the slepton masses.  This
implies suppression of the EDMs of the electron and muon, whereas CP
violation may occur in full strength in charged LFV processes, such as
$\mu\to eee$~\footnote{In the light of very stringent constraints from
electron, neutron and mercury EDMs~\cite{nath,khalil,masiero}, we neglect the
possible phases in diagonal soft supersymmetry-breaking terms throughout
this paper. This is natural in mechanisms which generate only real soft
terms, such as gravity-~\cite{grmed}, gauge-~\cite{gaugemed},  
anomaly-~\cite{ams}, gaugino-~\cite{ginom} and
radion-mediation~\cite{radion} mechanisms.}.

Another arena to probe LFV is provided by rare $\tau$ decays. There has
been some discussion in the literature of $\tau \to \ell \gamma$
decays~\cite{taumug}, and one could in principle hope to measure
CP-violating asymmetries in the various $\tau \to 3 \ell$ decays. Another
possibility is to search for LFV in sparticle decays \cite{acfh,sleptonosc}, 
e.g., $\tilde{\chi}^0_i \to e
\mu \tilde{\chi}^0_j$, where CP-violating asymmetries analogous to
$A_T(\mu\to eee)$ can also be defined in principle. However, we do not
investigate these possibilities further in this paper.

We concentrate here on CP-violating observables in the $\mu$
sector, assuming that the only sources of LFV and CP violation are the
interactions with heavy singlet $N^c$ neutrinos. We start by discussing
general parametrisations of the Yukawa matrix ${ Y_\nu}$ in terms of the
high- as well as the low-energy observables, paying particular attention
to the counting of physical degrees of freedom and their relations to 
CP-violating observables. Subsequently, we analyse
the renormalization-group running of soft supersymmetry-breaking terms,
assuming universal
boundary conditions at the GUT scale. 
Our first objective in this analysis is to
demonstrate in principle the complementarity of the different observables,
to see how all the CP-violating phases of the minimal seesaw model come
into play, and to clarify the relationship of the observable phases to the
phases appearing in leptogenesis. In the leading-logarithmic 
approximation, in which the heavy singlet neutrinos are treated as 
degenerate, this renormalization is sensitive to just one combination of 
the leptogenesis and light-neutrino phases, but two other combinations 
contribute beyond this 
approximation. We illustrate our results in a simple two-generation 
model. We then present numerical estimates of
${Br}(\mu\to e\gamma)$, ${Br}(\mu\to eee)$ and $A_T(\mu\to eee)$, 
taking into account the present knowledge
of neutrino mixings and masses as well as bounds on sparticle masses. 
We find that the magnitude of $A_T$ is in general
anti-correlated with the rate of $\mu\to e\gamma$, and may be large in 
some models compatible with the experimental upper limit on $\mu\to 
e\gamma$ decay. If a
cancellation occurs between different contributions to the $\mu-e-\gamma$
vertex, so that the box and penguin diagrams contributing to $\mu\to eee$
become comparable in magnitude, the T-odd asymmetry $A_T$ in $\mu\to eee$
may be as large as $\sim 10\%$, while ${Br}(\mu\to eee)$ remains
appreciable. However, the EDMs of the $\mu$ and $e$ are rather small in 
the minimal seesaw model.

It is important to note that the neutrino-oscillation phase $\delta$ and
the Majorana phases $\phi_{1,2}$ are completely independent of the three
physical phases in the quantity ${ Y_\nu Y_\nu^\dagger}$ that enters in
leptogenesis calculations.  On the other hand, $A_T$ and the other
renormalization-induced observables depend on mixtures of the 
light-neutrino
and leptogenesis phases. Thus, neutrino factories and LFV measurements
provide complementary information on the leptonic CP-violating phases.  
In particular, observation of $A_T$ is possible even if CP violation in
neutrino oscillations is unobservable, i.e., if either $\delta=0,$
$U_{e3}=0$ or the solar-neutrino deficit is not explained by the LMA
solution. However, in the minimal supersymmetric seesaw model, it is
possible that a combination of CP-violating observables in the neutrino
and charged-lepton sectors may provide constraints on the angles and
phases responsible for leptogenesis.

Our work is organized as follows. In Section 2, we consider general
parameterisations of the neutrino Yukawa couplings $Y_\nu$ and discuss CP
violation in the minimal supersymmetric seesaw model. In Section 3, we
give general formulae for the EDMs of the charged leptons, $\mu\to
e\gamma$, and $\mu\to eee$, including the latter's T-odd asymmetry $A_T$.
We present the results of the numerical analysis in Section 4. Finally,
Section 5 is devoted to a discussion and our conclusions concerning the
observability of the CP-violating phases in the minimal supersymmetric
seesaw model.


\section{CP Violation in the Lepton Sector of the Minimal Supersymmetric
Seesaw Model}

We consider the MSSM with three additional heavy singlet-neutrino
superfields ${N^c}_i$, constituting the minimal supersymmetric seesaw
model. The relevant leptonic part of its superpotential is
\begin{eqnarray}
\label{w}
W = N^{c}_i (Y_\nu)_{ij} L_j H_2
  -  E^{c}_i (Y_e)_{ij}  L_j H_1 
  + \frac{1}{2}{N^c}_i {\cal M}_{ij} N^c_j + \mu H_2 H_1 \,,
\label{MseesawM}
\end{eqnarray}
where the indices $i,j$ run over three generations and ${\cal M}_{ij}$
is the heavy singlet-neutrino mass matrix. Taking account of the
possible field redefinitions, this minimal supersymmetric seesaw model
contains 21 parameters: 3 charged-lepton masses $m_\ell$, 3 
light-neutrino masses ${\cal M}^D_\nu$, 3 heavy Majorana neutrino masses
${\cal M}^D$, 3 light-neutrino mixing angles $\theta_{ij}: 1 \le i \ne
j \le 3$, 3 CP-violating light-neutrino mixing phases $\delta,
\phi_{1,2}$ (the MNS phase and two Majorana phases),
and 3 additional mixing angles and 3 more phases associated with the
heavy-neutrino sector.

\subsection{High-Energy Parametrization}

In order to clarify the appearance and r\^oles of these parameters, we
first analyze (\ref{MseesawM}) in a basis where the charged leptons and
the heavy neutrinos both have real and diagonal mass matrices:
\begin{eqnarray}
(Y_e)_{ij} &=& {Y}^D_{e_i} \delta_{ij},
\nonumber\\
{\cal M}_{ij} &=& {\cal M}^D_i \delta_{ij},
\label{diagY}
\end{eqnarray}
where ${\cal M}^D=\mrm{diag}(M_{N_1},M_{N_2},M_{N_3})$.
{\it A priori}, the neutrino Yukawa-coupling matrix
${Y}_{\nu}$ has nine phases, which can be exposed by writing it in the
form: $Y_\nu = Z^\star {Y}^D_{\nu_k} X^\dagger\tilde{P}^\star_1,$ where
${Y}^D_{\nu}$ is diagonal and 
$\tilde{P}_1=\mrm{diag}(e^{i\sigma_{1}}, e^{i\sigma_{2}},
e^{i\sigma_{3}} )$. However, in the basis (\ref{diagY}) 
one may redefine the left-handed lepton fields
$L_i$, and thus rotate away the three phases in $\tilde{P}_1$, which are
unphysical. Thus the Yukawa-coupling matrix may be written in the form:
\begin{equation}
(Y_\nu)_{ij} = Z_{ik}^\star {Y}^D_{\nu_k} X_{kj}^\dagger.
\label{paramY}
\end{equation}
The matrix $X$ is the analogue in the lepton sector of the quark CKM
matrix, and thus it has only one physical phase. On the other
hand, we can always write $Z$ in the form
\begin{equation}
Z= P_1 \overline{Z} P_2,
\label{decomposeZ}
\end{equation}
where $\overline{Z}$ is a CKM-type matrix with three real mixing
angles and one physical phase, and
$P_{1,2}=\mrm{diag}(e^{i\theta_{1,3}}, e^{i\theta_{2,4}}, 1 )$ are
diagonal matrices containing two phases each. Thus $Z$ has 5 physical
phases to add to that in $X$, and all six real mixing angles and six
phase parameters in this basis are physical observables. 

We now study the combination $Y_\nu Y_\nu^\dagger$ of the Yukawa
couplings, which governs leptogenesis in this minimal seesaw model.
It is straightforward to see from (\ref{paramY}) that
\begin {equation}
\label{yy+1}
Y_\nu Y_\nu^\dagger 
 = P_1^\star \overline{Z}^\star (Y_\nu^D)^2 \overline{Z}^T P_1,
\end{equation}
which depends on just three of the CP-violating phases, namely the two
phases $\theta_{1,2}$ in $P_1$ and the single residual phase in
$\overline{Z}$, as well as the three real mixing angles in $\overline{Z}$.
This is consistent with the observation that, since the overall lepton
number involves a sum over the
light-lepton species (both charged leptons and light neutrinos), one would
not expect leptogenesis to depend on the 6 MNS angles and phases.

On the other hand, as we discuss in more detail below, mixing and CP
violation in the slepton sector of this minimal supersymmetric seesaw
model is controlled by the combination $Y_\nu^\dagger Y_\nu$ of the
neutrino Yukawa couplings, in the leading-logarithmic approximation where
$M_{GUT} \gg M_{N_{1,2,3}}$. It is again straightforward to see from
(\ref{paramY}) that
\begin{equation}
\label{y+y1}
Y_\nu^{\dagger} Y_{\nu} = X (Y_\nu{^D})^2 X^{\dagger} .
\end{equation}
Therefore, in this approximation, CP violation in charged LFV processes
arises only from the one physical phase in the diagonalizing matrix $X$.

\subsection{Low-Energy Parametrization}

We now reconsider leptonic CP violation from a more familiar point of
view \cite{valle}, 
namely that of the effective low-energy theory obtained after the
heavy neutrinos are decoupled. In this energy range, physics is
described by the following effective superpotential:
\begin{eqnarray}
\label{weff}
W_{eff}&=& L_i H_2 \left(Y_\nu^T \left({\cal M}^{D}\right)^{-1} 
Y_\nu\right)_{ij} L_j H_2 
  -  E^{c}_i (Y_e)_{ij}  L_j H_1 ,
\label{LEET}
\end{eqnarray}
where the effective light-neutrino masses are given in the basis
(\ref{diagY}) by
\bea 
{\cal M}_\nu={Y}_\nu^T \left({\cal M}^{D}\right)^{-1} {
Y}_\nu v^2 \sin^2\beta,
\label{seesaw1}
\eea
where $v=174$ GeV and as usual $\tan\beta=v_2/v_1$.
The mass matrix ${\cal M}_\nu$ can be diagonalized by a unitary matrix 
$U$:
\bea
U^T {\cal M}_\nu U = {\cal M}^D_\nu\,,
\label{Mnud}
\eea
where ${\cal M}^D_\nu=\mrm{diag}(m_{\nu_1},m_{\nu_2},m_{\nu_3}).$
Since ${\cal M}_\nu$ is a symmetric matrix and contains in general
six phases, $U$ must also have 6 phases. It can be expressed in the form
\begin{equation}
U=\tilde{P}_2 V P_0,
\label{generalU}
\end{equation}
where $P_0=\mrm{diag}(e^{-i\phi_1}, e^{-i\phi_2}, 1 ),$
$\tilde{P}_2=\mrm{diag}(e^{i\alpha_{1}}, e^{i\alpha_{2}}, 
e^{i\alpha_{3}} )$ and $V$ is the MNS matrix written in
the CKM form:
\bea 
\label{V} 
V=\pmatrix{c_{13}c_{12} & c_{13}s_{12} & s_{13}e^{-i\delta}\cr
-c_{23}s_{12}-s_{23}s_{13}c_{12}e^{i\delta} & 
c_{23}c_{12}-s_{23}s_{13}s_{12}e^{i\delta} & s_{23}c_{13}\cr
s_{23}s_{12}-c_{23}s_{13}c_{12}e^{i\delta} & 
-s_{23}c_{12}-c_{23}s_{13}s_{12}e^{i\delta} & c_{23}c_{13}\cr}.  
\eea
The phases in $\tilde{P}_2$ (\ref{generalU}) can be removed by
redefinition of the $L_i$ fields, leading to a new basis in which 
\bea
U=V P_0.
\label{U}
\eea 
This differs from the basis (\ref{diagY}) by the phase rotation
$\tilde{P}_2$. The new basis is appropriate if one works with the
effective low-energy observables in the effective superpotential
(\ref{weff}), e.g., for studying neutrino oscillations. Indeed, the
mixing angles $\theta_{ij}: 1 \le i \ne j \le 3$, whose $\sin, \cos$
we denote by $s_{ij},c_{ij}$, are measurable in neutrino-oscillation
experiments, as is the CP-violating MNS phase $\delta$. One
combination of two CP-violating Majorana phases $\phi_{1,2}$ is in
principle measurable in $\beta \beta_{0 \nu}$ experiments.

The physical interpretation of the Yukawa couplings in (\ref{w}) is made
more transparent in the basis (\ref{diagY}), which does not contain the
unphysical low-energy phases in $\tilde{P}_2$ that we rotated away in the
previous paragraph. Note that one must change $X\to \tilde{P}_2 X$ if one
works in the basis (\ref{U}). 

Our objective in this paper is to study CP-violating observables which are
sensitive to different physical phases. For this purpose, we need a proper
parametrization of the input parameters of the model.  The most
straightforward choice is to work in the basis (\ref{diagY}) and to choose
the physical observables in (\ref{diagY}) and (\ref{paramY}) 
as the input parameters.  In
this case, the physics is entirely transparent. However, the present
experiments do not measure heavy-neutrino masses, their Yukawa-coupling
and mixings directly. All the information we have on neutrinos comes from
the low-energy neutrino-oscillation and $\beta \beta_{0 \nu}$ experiments. 
If we choose the input parameters from (\ref{diagY}), (\ref{paramY}),
we have to check
every time that the induced ${\cal M}_\nu$ in (\ref{seesaw1})  agrees with
the experimental data. Instead, one can attempt to use the effective
low-energy observables as an input. 

To this end, we first rewrite the seesaw mechanism in the different form:
\beq
\label{seesaw4}
R \equiv \sqrt{{\cal M}^{D}}^{-1}Y_\nu U \sqrt{{\cal M}_\nu^{D}}^{-1} \; :
\; R^T R=1,
\eeq 
which is equivalent to (\ref{seesaw1}).
Starting with any given $Y_\nu$ and ${\cal M}^D$ as input
parameters, we obtain as outputs the seesaw-induced low-energy parameters
${\cal M}_\nu^{D}$ and $U$, and an auxiliary complex orthogonal
matrix $R$. It is possible to choose different
parameter sets for $Y_\nu$ and ${\cal M}^D$ that give the same low
energy effective ${\cal M}_\nu^{D}$ and $U$, but lead to different values
for $R$. 

One can turn the argument around~\cite{ci}, and parameterize 
the neutrino Yukawa-coupling matrix in 
terms of an arbitrary complex orthogonal matrix $R'$ as follows:
\bea { Y_\nu'}= \frac{\sqrt{{\cal
M}^D} R' \sqrt{{\cal M}^D_\nu}\, U^\dagger}{v\sin\beta}.
\label{Ynu}
\eea
We emphasize that the output $Y_\nu'$ in this parametrization is in
the {\it low-energy} basis (\ref{U}), and therefore contains {\it
unphysical} phases. If one wants to use the induced $Y_\nu'$ to
parametrize the superpotential (\ref{w}), one should be careful to
count correctly the physical degrees of freedom. 

We now form the combinations $Y_\nu' Y_\nu^{'\dagger}$ and 
$Y_\nu^{'\dagger} Y_\nu'$ out of (\ref{Ynu}). In the first case, we obtain
\begin{eqnarray}
Y_\nu' Y_\nu^{'\dagger} 
=\frac{\sqrt{{\cal M}^D} R' {\cal M}_\nu^D R^{'\dagger} \sqrt{{\cal M}^D}}
              {v^2 \sin^2\beta},
\label{yy+2}
\end{eqnarray}
which contains three independent physical phases that are
given entirely 
in terms of the parameters in the orthogonal matrix $R'.$ 
This is consistent with (\ref{yy+1}), and the new parametrization
therefore has not changed the counting of phases in $Y_\nu
Y_\nu^{\dagger}.$ On the other hand, we also obtain from (\ref{Ynu})
\begin{eqnarray}
Y_\nu^{'\dagger} Y_\nu' 
= U \frac{\sqrt{{\cal M}_\nu^D} R^{'\dagger} 
{\cal M}^D R' \sqrt{{\cal M}^D_\nu}}
              {v^2 \sin^2\beta} U^\dagger .
\label{y+y2}
\end{eqnarray}
This expression also appears to contain three phases, which are
combinations of all the parameters in $U$ and $R'$. 

However, according to (\ref{y+y1}), $Y_\nu Y_\nu^{\dagger}$ is supposed to
contain only one physical phase. What has happened? The answer is that
physics has not changed, and thus two out of three phases in
$Y_\nu^{'\dagger} Y_\nu'$ are unphysical. This is the case because we are
working in the low-energy basis (\ref{U}), and not in the basis
(\ref{diagY}). The three phases in $\tilde{P}_2$, which were rotated away
in defining $U$, appear now in $Y_\nu'$. Instead of (\ref{y+y1}), we now
have $Y_\nu^{'\dagger} Y'_\nu= \tilde{P}_2 X(Y_\nu^D)^2
X^\dagger\tilde{P}_2^\star$. One overall phase is irrelevant, and the two
unphysical relative phases in $\tilde{P}_2$ explain the faulty phase
counting in (\ref{y+y2}). 

In the following, we show explicitly that the unphysical phases in
$\tilde{P}_2$ cancel out in the Jarlskog invariants which can be
constructed using $Y_\nu^{'\dagger} Y_\nu'$.  Therefore, in the
leading-logarithmic approximation, all the CP-violating LFV observables
depend only on the one physical phase in (\ref{y+y2}), which is a
combination of the phases in $U$ and $R'.$ {\it Henceforward, we omit the
superscript $'$, but one must still be careful to distinguish between the
different bases.}

\subsection{Relations to CP-Violating Observables}

So far we have only considered the parametrization of the input neutrino
parameters, which in general are complex, and the 6 resulting independent
CP-violating phases. We now consider how physical observables depend on
these various phases. 

\subsubsection{Leptogenesis}

At present, our only experimental knowledge on CP violation in the lepton
sector may be obtained from the baryon asymmetry of the Universe, assuming
that this originated from leptogenesis. In leptogenesis scenarios,
initial $B-L$ asymmetries $\varepsilon^i$ appeared in decays of the heavy
neutrinos $N^c_i$ in the early Universe, as results of interferences
between the tree-level and one-loop amplitudes for $N^c_i$ decays. The $L$
asymmetry in the decay of an individual species $N^c_i$ is given in the
supersymmetric case~\cite{vissani} by
\begin{eqnarray}
\varepsilon^i &=& -\frac{1}{8 \pi} \sum_{l} 
\frac{ \mbox{Im}\Big[
\left( { Y_\nu}{ Y_\nu}^\dagger  \right)^{li}
\left( { Y_\nu}{ Y_\nu}^\dagger \right)^{li}
\Big]}
{ \sum_{j} |{ Y_\nu}^{ij}|^2 }
\nn \\
& &
\sqrt{x_l} \Big[  \mbox{Log} (1+1/x_l) +  \frac{2}{(x_l-1)}\Big] , 
\label{eps}
\end{eqnarray}
where $x_l=(M_{N_l} / M_{N_i})^2$ and both triangular and self-energy
type loop diagrams are taken into account. This $L$ asymmetry is
converted into the observed baryon asymmetry by sphalerons acting
before the electroweak phase transition. It is clear from (\ref{eps})
that the generated asymmetry depends only on the phases in ${ Y_\nu}{
Y_\nu}^\dagger$. Hence, according to the parametrization (\ref{Ynu}),
the only phases entering in the calculation of the baryon asymmetry of
the Universe are those in $R$. In order to demonstrate the feasibility
of leptogenesis, it would be necessary to prove that at least one of
the phases in $R$ is non-zero. Moreover, as we shall see, at least one
of the real part of the mixing angles in $R$ must also be non-zero,
and one would need to control other parameters, such as the
heavy-neutrino mass spectrum, before being able to calculate the
baryon asymmetry in terms of $R$, or vice versa.

\subsubsection{CP Violation in Neutrino Oscillations}

Measuring this is one of the main motivations for building neutrino
factories. We assume that the real MNS mixing angles $\theta_{ij}$ and
the mass-squared differences $\delta m_{ij}^2 \equiv m_{\nu_i}^2 -
m_{\nu_j}^2$ are all non-vanishing, in which case the the MNS phase
$\delta$ in (\ref{V}) is in principle observable.  It is,
realistically, observable in long-baseline neutrino factory
experiments if the LMA solution of the solar
neutrino problem is correct. The Majorana phases $\phi_{1,2}$ do not
affect neutrino oscillations at observable energies, but do affect
$\beta \beta_{0 \nu}$ decay. The conventional nuclear $\beta \beta_{0
\nu}$ experiments measure one combination of the light-neutrino masses
$m_{\nu_i}$ and the Majorana phases $\phi_{1,2}$. As in the CKM case,
one can introduce a Jarlskog invariant that characterizes the strength
of CP violation in neutrino oscillations:
\bea
J_{\nu}&=&
\mrm{Im}
\left[ 
\left( {\cal M}_\nu^\dagger {\cal M}_\nu  \right)_{12}
\left( {\cal M}_\nu^\dagger {\cal M}_\nu  \right)_{23}
\left( {\cal M}_\nu^\dagger {\cal M}_\nu  \right)_{31}
\right]\,
\nonumber\\
&=&
 \delta m^2_{12}
~\delta m^2_{23}
~\delta m^2_{31}
~\mrm{Im}
\left[ 
V_{11}
V_{12}^\star
V_{22}
V_{21}^\star
\right] .
\label{Jnu}
\eea
%
One sees explicitly that the Majorana phases $\phi_{1,2}$
cancel out in $J_{\nu}$.

It is clear from (\ref{seesaw1}) that, from the high-energy point of view,
$\delta$ depends on all the six independent phases in $Y_\nu$, including
those in the combinations $ Y_\nu Y_\nu^{\dagger}$ and $Y_\nu^{\dagger}
Y_\nu.$ On the other hand, in the low-energy parametrization of
(\ref{Ynu}), the phase $\delta$ is taken as an input parameter.

\subsubsection{Renormalization of Soft Supersymmetry-Breaking Terms:\\
Flavor-Changing Processes}

In the minimal supersymmetric seesaw model, renormalization induces
sensitivity to the neutrino Yukawa couplings ${ Y}_\nu$ in the soft
supersymmetry-breaking parameters in the slepton sector, in particular
to the CP-violating phases in ${ Y}_\nu$. These may have measurable
effects on several CP-violating lepton observables, including
asymmetries in LFV decays, which are
observable in rare $\mu$ and/or $\tau$ decays, 
and electric dipole moments. The $\mu$
electric dipole moment as well as rare $\mu$ decays may be measurable
using slow or stopped muons produced at the front end of a neutrino
factory. In this subsection, we concentrate on the flavor-changing
processes, such as asymmetries in LFV decays, and we will discuss
flavor-conserving processes, such as the electric dipole moment.

The soft supersymmetry-breaking terms in the leptonic sector of the
minimal supersymmetric seesaw model are
\begin{eqnarray}
-{\cal L}_{\rm soft}&=& 
  \tilde{L}_{i}^{\dagger} (m_{\tilde L}^2)_{ij}\tilde{L}_{j} 
+ \tilde{ E}_{i}^{c\ast} (m_{\tilde E}^2)_{ij}\tilde{E}_{j}^{c}  
+ \tilde{N}_{i}^{c\ast} (m_{{\tilde N}}^2)_{ij} \tilde{N}^{c}_{j}  
\nonumber\\
&& 
 + \left(\tilde{N}^{c}_{i} (A_{N})_{ij} \tilde{L}_{j} H_2
  -\tilde{E}^{c}_{i} (A_{e})_{ij}\tilde{L}_{j} H_1
 +\frac12 \tilde{N}_{i}^{c\ast} (B_{N})_{ij}\tilde{N}^{c}_{j}  
\right. \nn \\
&& \left.
 +\frac12 M_1 \tilde{B} \tilde{B}
 + \frac12 M_2 \tilde{W}^a \tilde{W}^a 
 + \frac12 M_3 \tilde{g}^a \tilde{g}^a
   + h.c.\right) .
\end{eqnarray}
We assume that the soft supersymmetry-breaking terms have
universal boundary conditions at the GUT scale $M_{GUT} \sim 2\times
10^{16}$ GeV:
\bea
& 
(m^2_{\tilde E})_{ij}=(m^2_{\tilde L})_{ij}=(m^2_{\tilde N})_{ij}=
m_0^2 {\bf 1},
&\nn \\
& m^2_{H_1}=m^2_{ H_2}=m_0\,, & \nn\\
&(A_e)_{ij}=A_0 (Y_e)_{ij}\,,(A_\nu)_{ij}=A_0 (Y_\nu)_{ij}\,,   & \nn\\
& M_1=M_2=M_3=m_{1/2}\,.&
\label{mssmbound}
\eea
At lower energies below $M_{GUT}$ and above the heavy-neutrino mass scale 
$M_N$, which we assume
to be $\ll M_{GUT}$, off-diagonal entries in ${ Y}_\nu$ generate via the
renormalization-group running
off-diagonal entries in the effective soft supersymmetry-breaking terms.
In the leading-logarithmic approximation the flavor-dependent parts of the 
soft supersymmetry-breaking terms are given by
\bea
\label{llrge} 
\left(\delta m_{\tilde L}^2\right)_{ij} & \approx & 
-\frac{1}{8\pi^2}(3m_0^2 + A_0^2)
({ Y_\nu^\dagger}{ Y_\nu}  +{ Y_e^\dagger}{ Y_e})_{ij}
\log\frac{M_{GUT}}{M_N}\ , 
\nn \\
\left(\delta m_{\tilde E}^2\right)_{ij} & \approx & 
-\frac{1}{4\pi^2}(3m_0^2 + A_0^2)
({ Y_e}{ Y_e^\dagger})_{ji}
\log\frac{M_{GUT}}{M_N}\ , 
\nn \\
(\delta {A_e})_{ij}& \approx &  
- \frac{1}{8\pi^2} A_0 Y_{e_i} (
3 { Y_e^\dagger}{Y_e}
+ { Y_\nu^\dagger}{Y_\nu})_{ij}
\log\frac{M_{GUT}}{M_N} .
\eea
Here, the Yukawa coupling constants are given at $M_N$, and then
${Y_e}$ is diagonal. This means that $m^2_{\tilde E}$ remains diagonal
in this approximation. Below $M_N$, the heavy neutrinos decouple, and
the renormalization-group running is given entirely in terms of the
MSSM particles and couplings, and is independent of $Y_\nu$. We use in
our numerical examples full numerical solutions to the one-loop
renormalization-group equations, but the approximate analytical
solutions (\ref{llrge}) are useful for a qualitative analysis.

It is important to notice that, in the leading-logarithmic
approximation (\ref{llrge}), the only combination of neutrino Yukawa
couplings entering the renormalization-group equations is
$Y_\nu^{\dagger} Y_\nu$. This implies that CP-violating phases are
induced only in the off-diagonal elements of $(m_{\tilde L}^2)_{ij}$
and $({A_e})_{ij}$, and further indicates that the lepton-flavour
conserving but CP-violating observables like the electric dipole
moments of charged leptons are naturally suppressed~\cite{khalil},
while CP violation in the charged LFV processes should occur in full
strength. This is analogous to CP violation in the quark sector of the
Standard Model, which is also directly related to flavour-changing
processes. As we saw earlier (\ref{y+y1}), the combination
$Y_\nu^{\dagger} Y_\nu$ depends on just one CP-violating phase, namely
that in the matrix $X$. Therefore, in the leading-logarithmic
approximation, {\it all} slepton-induced observables are independent
of the phases associated with leptogenesis, which are combinations of
those in the matrices $\overline{Z}$ and $P_1$ (\ref{y+y1}), in the
high energy parametrization. On the other hand, in the low-energy 
parametrization, 
$Y_\nu^{\dagger} Y_\nu$ depends on one combination of the phases in 
$U$ and $R$, as explained in subsection 2.2.

Since $Y_\nu^{\dagger}Y_\nu$ depends on only one physical phase, there is
only one invariant
for $m_{\tilde L}^2$  describing the strength of CP violation in any
process induced by sleptons. By analogy with the Standard Model quark
sector, this can be taken to be~\cite{acfh}
\bea 
J_{\tilde L}=\mrm{Im} \left[ \left(m_{\tilde
L}^2\right)_{12} \left(m_{\tilde L}^2\right)_{23} \left(m_{\tilde
L}^2\right)_{31} \right].  
\label{Jl}
\eea 
Additional invariants including the $A$
terms can be constructed:
\bea
J_{A_{12}}=\mrm{Im} \left[ \left(A_e\right)_{12} \left(m_{\tilde
L}^2\right)_{23} \left(m_{\tilde
L}^2\right)_{31} \right]
\label{Ajl}
\eea
and cyclic permutations, and similar invariants with two or three
$\left(A_e\right)_{ij}$ factors. However,
in this model they are all
related to the basic invariant (\ref{Jl}), and proportional to
\begin{eqnarray}
\mrm{Im}\left[
(Y_\nu^{\dagger}Y_\nu)^{12}
(Y_\nu^{\dagger}Y_\nu)^{23}
(Y_\nu^{\dagger}Y_\nu)^{31}
\right]
&=&
 \delta Y^2_{\nu 12} 
~\delta Y^2_{\nu 23} 
~\delta Y^2_{\nu 31} 
~\mrm{Im}
\left[ 
X_{11}
X_{12}^\star
X_{22}
X_{21}^\star
\right] 
\end{eqnarray}
in the leading-logarithmic approximation (\ref{llrge}).
Here, $\delta Y^2_{\nu ij} \equiv (Y_{\nu_i}^D)^2-(Y_{\nu_j}^D)^2$.

The above analysis is modified when one includes in the
renormalization-group running effects associated with the non-degeneracy
of the heavy neutrinos: $M_{N_i} \ne M_{N_j}$. In this case, $(\delta
m^2_{\tilde L})_{ij}$ in (\ref{llrge}) is replaced as follows:
$(\delta m^2_{\tilde L})_{ij} \to (\delta
m^2_{\tilde L})_{ij} +
\left({\tilde \delta} m_{\tilde L}^2\right)_{ij}$, and
$\left({\tilde \delta} m_{\tilde L}^2\right)_{ij}$ is given by 
\bea
\label{nondegrge}
\left({\tilde \delta}
m_{\tilde L}^2\right)_{ij} \approx
-\frac{1}{8\pi^2}(3m_0^2 + A_0^2)
({ Y_\nu^\dagger} L { Y_\nu})_{ij}.: 
L \equiv \log\frac{M_{N}}{M_{N_i}} \delta_{ij}\ ,
\eea
where $M_{N}$ is now interpreted as the geometric mean of the heavy
singlet-neutrino mass eigenvalues $M_{N_i}$. The first term in
(\ref{nondegrge}) contains the matrix factor
\begin{equation}
Y^\dagger L Y = X Y^D P_2 {\overline Z}^T L {\overline Z}^*
P_2^* Y^D X^\dagger,
\end{equation}
which induces some dependence on phases in ${\overline Z} P_2$. In the
three-generation case, there are two independent entries in the traceless
diagonal matrix $L$, so the renormalization induces in principle
dependences on two new combinations of these phases, as well as the single
phase in ${Y_\nu^\dagger}{Y_\nu}$. Thus low-energy observables 
become sensitive to all three leptogenesis phases. However, the 
dependences on the two
extra phases are suppressed to the extent that $\log\frac{M_{N}}{M_{N_i}} \ll
\log\frac{M_{GUT}}{M_N}$. 

\subsubsection{Renormalization of Soft Supersymmetry-Breaking Terms:\\
Flavor-Conserving Processes}

As mentioned above, since the CP-violating phases are in the
off-diagonal components of the soft supersymmetry-breaking terms, the
electric dipole moment of lepton is naturally suppressed. The following is
the lowest-order combination of the Yukawa couplings $Y_{\nu}$ and
$Y_{e}$ whose diagonal components have imaginary parts:
\begin{eqnarray}
J^{(i)}_{\rm edm} &=& \mrm{Im} \left[\left[Y_e Y_\nu^\dagger Y_\nu 
\left[Y_e^\dagger Y_e,~ Y_\nu^\dagger Y_\nu \right] 
Y_\nu^\dagger Y_\nu \right]^{ii}\right] \,
\nonumber\\
&=&
2 Y_{e_i}
~\delta Y_{e jk}^2
~\delta Y_{\nu 12}^2
~\delta Y_{\nu 23}^2
~\delta Y_{\nu 31}^2
~\mrm{Im}
\left[ 
X_{11}
X_{12}^\star
X_{22}
X_{21}^\star
\right] \sum_k \epsilon_{ijk} 
\label{jedm}
\end{eqnarray}
and the dominant contributions to the electric dipole moment are
proportional to it\footnote{
Similar studies for the electric dipole moment of neutron in the
MSSM are done in \cite{nedm}, assuming that  all CP violating phases  
come from the CKM matrix. 
}.
 Since $Y_\nu^\dagger Y_\nu$ in $J^{(i)}_{\rm edm}$
comes from the radiative correction to the soft supersymmetry-breaking
terms, the leading contribution to the electric dipole moment is
proportional to $\log^3\frac{M_{GUT}}{M_N}$ when
$\log\frac{M_{GUT}}{M_N}\ll 4\pi$. The dependence on $Y_e^\dagger
Y_e$ in $J^{(i)}_{\rm edm}$ comes from the radiative correction in the
soft supersymmetry-breaking terms or the tree-level mass matrix of the
charged sleptons. In this subsection, we present the Jarlskog
invariant for the soft supersymmetry-breaking terms contributing to 
the
electric dipole moment in ${\cal O}(\log^3 \frac{M_{GUT}}{M_N})$.
Also, we discuss cases where this approximation is invalid, namely 
when {\it
i)} $\tan\beta \gg 1$, or {\it ii)} 
non-degeneracy
between the heavy singlet neutrinos induces dependences of $m_{\tilde
L}^2$ and $A_e$ on phases in the product ${\overline Z} P_2$.

In order to evaluate the contribution to the electric dipole moment 
in ${\cal O}(\log^3\frac{M_{GUT}}{M_N})$, we need the corrections to the 
soft supersymmetry-breaking terms at ${\cal O}(\log^2 
\frac{M_{GUT}}{M_N})$, which are
\bea
\label{llrge2} 
\left(\delta^{(2)} m_{\tilde L}^2\right)_{ij} & \approx & 
\frac{4}{(4\pi)^2} A_0^2
(3{ Y_\nu^\dagger}{ Y_\nu} { Y_\nu^\dagger}{ Y_\nu} 
 +
3  { Y_e^\dagger}{ Y_e}{ Y_e^\dagger}{ Y_e}
+\{{ Y_e^\dagger}{ Y_e},~{ Y_\nu^\dagger}{ Y_\nu}\})_{ij}
\log^2\frac{M_{GUT}}{M_N}\ , 
\nn \\
\left(\delta^{(2)} m_{\tilde E}^2\right)_{ij} & \approx & 
\frac{8}{(4\pi)^2} A_0^2
(3 { Y_e}{ Y_e^\dagger}{ Y_e} { Y_e^\dagger}
+{ Y_e}{ Y_\nu^\dagger}{ Y_\nu}{ Y_e^\dagger} )_{ji}
\log^2\frac{M_{GUT}}{M_N}\ , 
\nn \\
(\delta^{(2)} {A_e})_{ij}& \approx &  0 .
\eea
Here, we neglect irrelevant terms with a trace over flavor indices, or 
which are flavor-independent. The Yukawa couplings are evaluated at $M_N$.
From these equations and (\ref{llrge}), non-vanishing contributions to
$J^{(i)}_{\rm edm}$ arise from the following combinations of
${\cal O}(\log^3 \frac{M_{GUT}}{M_N})$:
\begin{eqnarray}
\mrm{Im} \left[\delta A_e \delta A_e^\dagger \delta A_e \right]_{ii}
&=&
\frac{4}{(4 \pi)^6} 
A_0^3 J_{\rm edm}^{(i)}
\log^3\frac{M_{GUT}}{M_N}.
\label{jedm1}
\label{jedm2p}
\\
\mrm{Im}\left[\delta {A_e} {Y_e^\dagger}{Y_e} 
\delta^{(2)} m_{\tilde L}^2\right]_{ii}
&=&
-\frac{12}{(4 \pi)^6} 
A_0^3 J_{\rm edm}^{(i)}
\log^3\frac{M_{GUT}}{M_N},
\label{jedm2}
\end{eqnarray}
In (\ref{jedm2}), the combination ${Y_e^\dagger}{ Y_e}$ arises from the 
tree-level mass matrix of the charged sleptons. It is found from
(\ref{jedm2p},\ref{jedm2}) that the electric dipole moments depend
strongly on $A_0$ and less on $m_0$.

When $\tan\beta \gg 1$, (\ref{jedm2}) is proportional to
$\tan^2\beta$ and (\ref{jedm1}) is not enhanced. On the other hand,
terms such as
\begin{eqnarray}
\mrm{Im}\left[\delta^{(2)}m_{\tilde E}^2 Y_{e} \delta^{(2)}
m_{\tilde L}^2\right],
\end{eqnarray}
are proportional to $\tan^3\beta$ and
$\log^4 \frac{M_{GUT}}{M_N}$. Thus, they may make sizeable
contributions to the electric dipole moments for $\tan\beta \gg 1$,
even if $\log \frac{M_{GUT}}{M_N}\lsim 4\pi$.

If the heavy neutrinos are not degenerate in mass, they induce
dependences of the soft supersymmetry-breaking terms on phases in
${\overline Z} P_2$, as mentioned in the previous Section, which then
contribute to the electric dipole moments. The Jarlskog invariant
$J^{(i)}_{\rm edm}$ depends on $Y_e^\dagger Y_e$, and this factor
suppresses the electric dipole moment when $\tan\beta$ is small. In
this case, the non-degeneracy of the heavy neutrinos may
have a more important effect on the electric dipole moment. The
corrections of ${\cal O}(\log \frac{M_{GUT}}{M_N}\log
 \frac{M_N}{M_{N_i}})$ are 
\bea
\left(\tilde{\delta}^{(2)} m_{\tilde L}^2\right)_{ij} & \approx & 
\frac{18}{(4\pi)^4} (m_0^2 +A_0^2)
\{
{ Y_\nu^\dagger}L { Y_\nu}, 
{Y_\nu^\dagger}{ Y_\nu} 
\}
\log\frac{M_{GUT}}{M_N}\ , 
\nn \\
\left(\tilde{\delta}^{(2)} m_{\tilde E}^2\right)_{ij} & \approx & 
0,
\nn \\
\left(\tilde{\delta}^{(2)} {A_e}\right)_{ij}& \approx &  
\frac{1}{(4\pi)^4} A_0 Y_e
(
11 \{
{ Y_\nu^\dagger}L { Y_\nu}, 
{Y_\nu^\dagger}{ Y_\nu} 
\}
+
7 [
{ Y_\nu^\dagger} L { Y_\nu}, 
{Y_\nu^\dagger}{ Y_\nu} 
]
)_{ij}
\log\frac{M_{GUT}}{M_N}\ . 
\eea
Here, we neglect terms with ${Y_e^\dagger}{ Y_e}$ factors. The 
interesting
point is that the second term in $\tilde{\delta}^{(2)} {A_e}$ can have
imaginary parts in the diagonal components, and thus can contribute to
the electric dipole moment~\footnote{
Whilst the combination $(\tilde{\delta}A_e \delta m_{\tilde
L}^2)_{ii}$ has an imaginary part, it does not contribute to the electric
dipole moments, since $\mrm{Im}[(\delta A_e + \tilde{\delta}A_e)
(\delta m_{\tilde L}^2+\tilde{\delta} m_{\tilde L}^2)]_{ii}=0$. 
}.
Since phases in $\tilde{\delta}^{(2)} {A_e}$ arise from
${\overline Z} P_2$, we do not need three generations of leptons in order 
for $\tilde{\delta}^{(2)} {A_e}$ to have imaginary parts in the diagonal
terms. The behaviour of this contribution will be discussed in 
the next subsection.

\subsection{Two-Generation Model}

We now demonstrate the interdependences of the above physical observables
in a toy two-generation model. In this model, $X$ has no physical phase
while there may be one phase in $Z$. We parametrize the 
light- and heavy-neutrino masses and $R$ as follows:
\begin{eqnarray}
&{\cal M}_\nu^D=
\left(
\begin{array}{cc}
m_{\nu_1} & 0 \\
0  & m_{\nu_2}
\end{array}
\right), 
\,\,\,
\,\,\,
\,\,\,
{\cal M}^D=
\left(
\begin{array}{cc}
M_{1} & 0 \\
0  & M_{2}
\end{array}
\right) ,& 
\nn \\
&
R=
\left(
\begin{array}{cc}
\cos(\theta_r+i \theta_i)  & \sin(\theta_r+i \theta_i)   \\
-\sin(\theta_r+i \theta_i)  & \cos(\theta_r+i \theta_i)   
\end{array}
\right).
&
\end{eqnarray}
In this model, the leptogenesis invariant is $\mbox{Im}\Big[ \left( {
Y_\nu}{ Y_\nu}^\dagger \right)^{21} \left( { Y_\nu}{ Y_\nu}^\dagger
\right)^{21} \Big] = 2 \times \mbox{Im}\Big[ \left( { Y_\nu}{
Y_\nu}^\dagger \right)^{12} \Big] \times \mbox{Re}\Big[ \left( { Y_\nu}{
Y_\nu}^\dagger \right)^{12} \Big]$, where
\begin{eqnarray}
\mbox{Im}\Big[ \left( { Y_\nu}{ Y_\nu}^\dagger  \right)^{12} \Big]
&=&
\frac{(m_{\nu_1}+m_{\nu_2}) \sqrt{M_1M_2}}{2 v^2\sin^2\beta}
{\rm sinh}2 \theta_i, \nn \\
\mbox{Re}\Big[ \left( { Y_\nu}{ Y_\nu}^\dagger  \right)^{12} \Big]
&=&
-\frac{(m_{\nu_1}-m_{\nu_2}) \sqrt{M_1M_2}}{2 v^2\sin^2\beta}
{\rm sin}2 \theta_r,
\label{ReIm}
\end{eqnarray}
so that
\begin{equation}
\mbox{Im}\Big[ \left( {
Y_\nu}{ Y_\nu}^\dagger \right)^{21} \left( { Y_\nu}{ Y_\nu}^\dagger
\right)^{21} \Big] = -\frac{(m^2_{\nu_1}-m^2_{\nu_2})M_1M_2}{2
v^4\sin^4\beta} {\rm sinh}2 \theta_i {\rm sin}2 \theta_r
\label{lepto2}
\end{equation}
As explained above, the phase $\theta_i$ in $R$ controls
leptogenesis, and the mixing angle $\theta_r$ must also be non-vanishing. 

As concerns neutrino observables, we recall that there is no analogue of
the MNS phase $\delta$ in this two-generation model. There is one
CP-violating Majorana phase $\phi$ for the light neutrinos, but this does
not contribute to leptogenesis, as we argued previously on general grounds
and now see explicitly in (\ref{lepto2}).

We now consider the quantity $Y_\nu^\dagger Y_\nu$ which controls the
renormalization of the soft super- symmetry-breaking terms in the
leading-logarithmic approximation, in particular $({Y_\nu}^\dagger {
Y_\nu})^{12}$ which has a non-zero imaginary part.  For illustrational
purposes, we assume that the light-neutrino mass matrix has maximal
mixing: 
\begin{eqnarray}
U &=&
\left(
\begin{array}{cc}
1/\sqrt{2} & 1/\sqrt{2}\\
-1/\sqrt{2} & 1/\sqrt{2}
\end{array}
\right) \left(
\begin{array}{cc}
{\rm e}^{-i \phi}  & 0\\
0 & 1
\end{array}
\label{U2}
\right)
\end{eqnarray}
where $\phi$ is a light-neutrino Majorana phase. In this case,
\begin{eqnarray}
\mbox{Re}\Big[\left( { Y_\nu}^\dagger { Y_\nu} \right)^{12} \Big]
&=&
-\frac{m_{\nu_2} + m_{\nu_1}}{4 v^2\sin\beta^2}
(M_1 - M_2) \cos 2 \theta_r 
\nonumber\\
&&
-
\frac{m_{\nu_2} - m_{\nu_1}}{4 v^2\sin\beta^2}
(M_1 + M_2) {\rm cosh} 2 \theta_i , 
\label{Rey+y} \\
\mbox{Im}\Big[\left( { Y_\nu}^\dagger { Y_\nu} \right)^{12} \Big]
&=&
\frac{\sqrt{m_{\nu_1}m_{\nu_2}}}{2 v^2\sin\beta^2}
(M_1+M_2) \cos\phi {\rm sinh} 2 \theta_i
\nonumber\\
&&
- 
\frac{\sqrt{m_{\nu_1}m_{\nu_2}}}{2 v^2\sin\beta^2}
(M_1-M_2) \sin\phi \sin 2 \theta_r .
\label{renorm2}
\end{eqnarray}
We see that the imaginary part of the off-diagonal component depends both
on the Majorana phase $\phi$ in $U$ (\ref{U2}) and the phase $\theta_i$ in
$R$. Even if it could be measured, and the neutrino mass eigenvalues
$M_{1,2}, m_{\nu_{1,2}}$ were known, still only one combination of the
angle factors $\theta_{r,i}$ entering in leptogenesis (\ref{lepto2}) would
be known, and there would still be an ambiguity associated with the
Majorana phase $\phi$. In fact, no CP violation is induced by the
renormalization (\ref{renorm2}) in this simple two-generation model, since
it is not possible to define the Jarlskog invariant $J_{\tilde L}$
(\ref{Jl}) and its analogues (\ref{Ajl}). Such invariants can be defined
in a three-generation model, and CP-violating observables are demonstrably
proportional to it, as we show in the next Section. 

As commented in subsections 2.3.3 and 2.3.4, non-degeneracy
between the heavy singlet neutrinos $N_i$ induces, via
renormalization, dependences of the entries of $m_{\tilde L}^2$ and
$A_e$ on phases in the product ${\overline Z} P_2$. In the
two-generation case, this dependence is on the one phase in $P_2$,
since ${\overline Z}$ has no phases in this case. This makes changes in
Arg$(m_{\tilde L}^2)_{12}$ and Arg$(A_e)_{12}$, but these are
suppressed to the extent that $\log\frac{M_{N}}{M_{N_i}} \ll
\log\frac{M_{GUT}}{M_N}$. Moreover, these small changes are identical
in the leading-logarithmic approximation.  On the other hand, the
corrections of the order of $\log\frac{M_{N}}{M_{N_i}}
\log\frac{M_{GUT}}{M_N}$ to the phases of the diagonal terms in $A_e$
may be sizeable. These are given in the two-generation model by
\begin{eqnarray}
\mrm{Im}\left[[
{ Y_\nu^\dagger} L { Y_\nu}, 
{Y_\nu^\dagger}{ Y_\nu} 
]^{11} \right]
&=&
\frac{\sqrt{m_{\nu_1} m_{\nu_2}}(m_{\nu_1}-m_{\nu_2}) M_1M_2}{2 v^4 \sin^4\beta} 
\log\frac{M_2}{M_1}
\cosh 2\theta_i \sin 2 \theta_r \sin\phi,
\nonumber\\
&+&
\frac{\sqrt{m_{\nu_1} m_{\nu_2}}(m_{\nu_1}+m_{\nu_2}) M_1M_2}{2 v^4 \sin^4\beta} 
\log\frac{M_2}{M_1}
\sinh 2\theta_i \cos 2 \theta_r \cos\phi
\nonumber\\
\mrm{Im}\left[[
{ Y_\nu^\dagger} L { Y_\nu}, 
{Y_\nu^\dagger}{ Y_\nu} 
]^{22} \right]
&=&
-\mrm{Im}\left[ [
{ Y_\nu^\dagger} L { Y_\nu}, 
{Y_\nu^\dagger}{ Y_\nu} 
]^{11}\right].
\label{39} 
\end{eqnarray}
We see that this effect vanishes if $M_2=M_1.$ 
Defining $M_2=M_1 (1+\delta),$ \Eq{39} 
grows with the dimensionless parameter $\delta$ (linearly, if $\delta$
is small) and is maximized when $\log[M_2 / M_1] = 1.$

\section{CP-Violating Observables in the Charged-lepton \\ Sector}

In this Section we discuss in more detail CP-violating and LFV observables
in the charged-lepton sector. The slepton-mixing effects discussed in the
previous Section generate LFV and CP-violating vertices involving
charginos, which in turn induce effective non-renormalizable interactions,
as we discuss in the following. 

\subsection{Chargino and Neutralino Interactions}

The relevant neutralino and chargino interactions
for leptons and sleptons are given by~\cite{h1,Okada}
\begin{eqnarray}
{\cal L}  & = & \overline{e_{i}}(N^{L}_{iAX}P_{L}+N^{R}_{iAX}P_{R})
                       \tilde{\chi}^{0}_{A}\tilde{e}_{X} 
           + \overline{e_{i}}(C^{L}_{iAX}P_{L}+C^{R}_{iAX}P_{R})
                       \tilde{\chi}^{-}_{A}\tilde{\nu}_{X} 
              +h.c.,
\end{eqnarray}
where $P_{R}=(1+\gamma_{5})/2,$ $P_{L}=(1-\gamma_{5})/2$
and
\begin{eqnarray}
N^{L}_{iAX} & = & -g\{ \sqrt{2}\tan\theta_{W}
                       (O_{N})^{}_{A1}(U_{e})_{X i+3}^{*}
                      +\frac{(m_{e})_{ij}}{\sqrt{2}m_{W}\cos\beta}
                       (O_{N})^{}_{A3}(U_{e})_{Xj}^{*} \},
 \\
N^{R}_{iAX} & = & -g[-\frac{1}{\sqrt{2}}
                       \{(O_{N})_{A2}^{*}
                   +\tan\theta_{W}(O_{N})_{A1}^{*} \}
                       (U_{e})_{Xi}^{*} \nonumber \\
            &   &  +\frac{(m_{e}^{\dag})_{ij}}{\sqrt{2}m_{W}\cos\beta}
                       (O_{N})_{A3}^{*}(U_{e})_{Xj+3}^{*} ], \\
C^{L}_{iAX} & = & g\frac{(m_{e})_{ij}}{\sqrt{2}m_{W}\cos\beta}
                  (O_{CL})_{A2}^{*}(U_{\nu})_{Xj}^{*},
 \\
C^{R}_{iAX} & = & -g(O_{CR})_{A1}^{*}(U_{\nu})_{Xi}^{*}.
\end{eqnarray}
In these expressions the matrices $O_N,$ $O_{CL},$ $O_{CR},$
$U_e$ and $U_\nu$ diagonalize the neutralino, left- and right-chargino,
charged-slepton and sneutrino mass matrices, respectively.
The indices run between $A=1,...,4$ for neutralinos, $A=1,2$ for charginos,
$X=1,...,6$ for sleptons and $X=1,2,3$ for sneutrinos.
In our framework the complex phases appear only in $U_e$ and $U_\nu.$

\subsection{LFV Muon Decays}

The effective Lagrangian for polarized
$\mu^{+} \rightarrow e^{+}\gamma$ 
and $\mu^{+} \rightarrow e^{+}e^{+}e^{-}$ decays is \cite{Okada}
\begin{eqnarray}
{\cal L} &=& -\frac{4G_F}{\sqrt{2}}\{  
        {m_{\mu }}{A_R}\overline{\mu_{R}}
        {{\sigma }^{\mu \nu}{e_L}{F_{\mu \nu}}}
       + {m_{\mu }}{A_L}\overline{\mu_{L}}
        {{\sigma }^{\mu \nu}{e_R}{F_{\mu \nu}}} \nonumber \\
    &&   +{g_1}(\overline{{{\mu }_R}}{e_L})
              (\overline{{e_R}}{e_L})
       + {g_2}(\overline{{{\mu }_L}}{e_R})
              (\overline{{e_L}}{e_R}) \nonumber \\
    &&   +{g_3}(\overline{{{\mu }_R}}{{\gamma }^{\mu }}{e_R})
              (\overline{{e_R}}{{\gamma }_{\mu }}{e_R})
       + {g_4}(\overline{{{\mu }_L}}{{\gamma }^{\mu }}{e_L})
              (\overline{{e_L}}{{\gamma }_{\mu }}{e_L})  \nonumber \\
    &&   +{g_5}(\overline{{{\mu }_R}}{{\gamma }^{\mu }}{e_R})
              (\overline{{e_L}}{{\gamma }_{\mu }}{e_L})
       + {g_6}(\overline{{{\mu }_L}}{{\gamma }^{\mu }}{e_L})
              (\overline{{e_R}}{{\gamma }_{\mu }}{e_R})
       +  h.c. \}.
\label{eq:effective}
\end{eqnarray}
Here $A_{L}$ and $A_{R}$ are the dimensionless 
photon-penguin couplings which 
induce $\mu^{+} \rightarrow e_{L}^{+}\gamma$ and 
$\mu^{+} \rightarrow e_{R}^{+}\gamma,$ respectively, which also contribute
to the $\mu^{+} \rightarrow e^{+}e^{+}e^{-}$ process, and the
$g_{i}: i=1,...,6$ are dimensionless four-fermion coupling constants 
which contribute only to $\mu^{+} \rightarrow e^{+}e^{+}e^{-}$.
Explicit expressions for $A_{L,R}$ and the $g_{i}$ in terms of 
$N_{iAX}^{L,R},$ $C_{iAX}^{L,R}$ are lengthy~\cite{Okada}, so we do not
rewrite them here.

In the notation (\ref{eq:effective}) the total $\mu^{+} \rightarrow
e^{+}\gamma$ branching ratio is given by 
\bea
{Br}(\mu^{+} \rightarrow e^{+}\gamma)=384 \pi^2 
\left(|A_L|^2+|A_R|^2 \right)\,,
\eea
and that of $\mu^{+} \rightarrow e^{+}e^{+}e^{-}$ by 
\bea
{{Br}}(\mu^+ \rightarrow e^+e^-e^+) &=& 2(C_1+C_2) +C_3+C_4+
32 \left\{\log\frac{m_\mu^2}{m_e^2}-\frac{11}{4}\right\} (C_5+C_6)
\nonumber \\
&&+16(C_7+C_8) + 8 (C_9+C_{10})\,.
\label{alltheCs}
\eea
The coefficients $C_{i}$ appearing in (\ref{alltheCs}) are functions of
$A_{L,R}$ and $g_{i}$:
\bea
&& C_{1} = \frac{|g_{1}|^{2}}{16} + |g_{3}|^{2},~  
C_{2} = \frac{|g_{2}|^{2}}{16} + |g_{4}|^{2},~ \nonumber \\
&& C_{3} = |g_{5}|^{2},~ C_{4} = |g_{6}|^{2},C_{5} = |eA_{R}|^{2},~   
C_{6}   =   |eA_{L}|^{2},~
 C_{7}   =   {\rm Re}(eA_{R}g_{4}^{*}), \nn \\
&&C_{8}   =   {\rm Re}(eA_{L}g_{3}^{*}),~
C_{9}   =   {\rm Re}(eA_{R}g_{6}^{*}),~  
C_{10}   =   {\rm Re}(eA_{L}g_{5}^{*})\,.
\eea
In order for CP violation to appear in any process, interference between
different terms in the amplitude for the process must occur. Therefore,
all possible observables in $\mu\to e\gamma$ decays, such as differences
between the $\mu^{+} \rightarrow e^{+}\gamma$ and $\mu^{-} \rightarrow
e^{-}\gamma$ rates, vanish in the leading order of perturbation theory. 
Moreover, the process $\mu^{-} \rightarrow e^{-}\gamma$ is not measurable
with high accuracy because of the large backgrounds. However, when muons
are polarized, a T-odd asymmetry for final-state particles in $\mu^{+}
\rightarrow e^{+}e^{+}e^{-}$ can be defined. Since CPT is conserved, the
T-odd asymmetry measures the amount of CP violation in our model. 

The muon polarization vector $\vec{P}$ can be defined in the coordinate
system in which the $z$ axis is taken to be the direction of the electron
momentum, the $x$ axis the direction of the most energetic positron
momentum, and the $(z\times x)$ plane defines the decay plane
perpendicular to the $y$ axis. It is necessary to introduce an energy
cutoff for the more energetic positron: $E_1< (m_\mu/2)(1-\delta).$ We use
$\delta=0.02$ to optimize the T-odd asymmetry, following~\cite{Okada}. 
Assuming $100\%$-polarized muons the T-odd asymmetry is then defined by
\bea
A_{T} &=& \frac{N(P_{\it y} >0)-N(P_{\it y} <0)}
{N(P_{\it y} >0)+N(P_{\it y} <0)} =
\frac{3}{2 {Br}(\delta=0.02)}
\left\{2.0 C_{11}-1.6 C_{12} \right\},
\label{AT}
\eea
where $N(P_i >(<)0)$ 
denotes the number of events with a positive (negative) $P_i$ 
component for the muon polarization,
\bea
C_{11}   =   {\rm Im} (eA_{R}g_{4}^{*}+eA_{L}g_{3}^{*}),~~~~
C_{12}   =   {\rm Im} (eA_{R}g_{6}^{*}+eA_{L}g_{5}^{*})\,,
\eea
and the branching ratio for $\delta=0.02$ is
\bea
{Br}(\delta=0.02) &=& 1.8 (C_1+C_2)+0.96 (C_3+C_4)
+88 (C_5+C_6)  \nonumber \\
&& +14 (C_7+C_8) +8 (C_9+C_{10}).
\label{brd}
\eea
It is known~\cite{Okada} that the asymmetry (\ref{AT}) may be large in
SU(5) SUSY GUTs~\cite{bhs}.  We study below whether this is also the case
in the minimal supersymmetric seesaw model. 

\subsection{Electric Dipole Moments}

The electric dipole moment of a generic lepton $\ell$ is defined as 
the coefficient $d_\ell$ of the interaction
\bea
{\cal L} = 
-\frac{i}{2} d_\ell\,\bar \ell\, \sigma_{\mu\nu}\gamma_5\, \ell\,
F^{\mu\nu}.
\label{edm_eff}
\eea
The current experimental bounds are $d_e<4.3\times 10^{-27}$ e cm for the
electron \cite{eedm}, $d_\mu = (3.7 \pm 3.4) \times 10^{-19}$ e cm for the muon \cite{muedm}, and
$|d_\tau| < 3.1
\times 10^{-16}$ e cm for the $\tau$ \cite{tauedm}. An experiment has been proposed at
BNL that could improve the sensitivity to $d_\mu$ down to $d_\mu \sim
10^{-24}$ e cm \cite{mnbuedm}, and PRISM and neutrino factory experiments aim at
sensitivities $d_\mu \sim 5\times 10^{-26}$ e cm. These bounds will impose
serious constraints on CP violation in the MSSM, as will prospective
improvements in the sensitivity to $d_e$ and $d_\tau$. 

In the MSSM, the $d_\ell$ receive contributions from chargino and
neutralino loops:
\bea
d_l=d^{\chi^+}_l+d^{\chi^0}_l\,,
\eea
where \cite{nath,khalil}
\bea
\label{dl+}
 d_l^{\chi^+}&=&-\frac{e}{(4\pi)^2} 
      \sum_{A=1}^{2}\sum_{X=1}^{3} 
 {\rm Im}(C^L_{lAX} C^{R*}_{lAX})\;   
 {m_{\chi^+_A}\over {m_{\tilde{\nu}_X}^2}}
   {\rm A}\biggl( \frac{m_{\chi^+_A}^2}{m_{\tilde{\nu}_X}^2} \biggr)\;,
\label{dlc}
\\ 
\label{dl0}
 d_l^{\chi^0 }&=&-\frac{e}{(4\pi)^2}
   \sum_{A=1}^{4}\sum_{X=1}^{6} 
{\rm Im}( N^L_{lAX} N^{R*}_{lAX}  )
               \frac{m_{\chi^0_A}}{M_{\tilde{l}_X}^2}\;
{\rm B}\biggl( \frac{m_{\chi^0_A}^2}{M_{\tilde{l}_X}^2}\biggr) \;,
\eea
and the loop functions are given by
\bea
&& A(r)=\frac{1}{2(1-r)^2}\biggl(3-r+\frac{2\log r}{1-r}\biggr) \;, 
\nonumber\\
&& B(r)=\frac{1}{2(r-1)^2}\biggl(1+r+\frac{2r\log r}{1-r}\biggr) \; . \nn
\eea
where the relevant chargino and neutralino couplings were given above.
We do not consider in our analysis the possibility of CP violation in the
chargino and neutralino mass matrices.

\subsection{R\^{o}le of the Jarlskog Invariants}

We first present approximate formulae for the effective couplings in
(\ref{eq:effective}), in order to show the qualitative behaviours of the
LFV processes and demonstrate the r\^{o}le of the Jarlskog invariants.
Since $(\delta m_{\tilde L}^2)$ and $(\delta {A_e})$ are the only sources
of off-diagonal components, the only non-negligible terms are $A_R$, $g_4$
and $g_6$: other terms are suppressed by the electron or muon masses. For
illustrative purposes in this subsection only, we assign to the soft
supersymmetry-breaking parameters common value $m_{S}$ at the
electroweak scale: 
\begin{eqnarray}
&M_2=M_1=\mu=(A_{e})_{22}/Y_{e_2}\equiv m_{S}& \, ,
\nonumber\\
&(m^2_{\tilde{L}})_{ii}=(m^2_{\tilde{E}})_{ii}\equiv m_{S}^2\, .&
\label{limit1}
\end{eqnarray}
Assuming $m_{S}\gg m_Z$, we then find
\begin{eqnarray}
&A_R = 4.8 \times 10^{-5} \left(\frac{\rm 100GeV}{m_{S}}\right)^2&
\nonumber\\
&
\times \left\{
\Delta_{21}^{\tilde{L}}
-0.64      \Delta_{23}^{\tilde{L}}\Delta_{31}^{\tilde{L}}
+0.66      \Delta_{21}^{A_e}
-0.40      \Delta^{A_e}_{23} \Delta^{\tilde{L}}_{31}
+\tan\beta
(
2.4 \Delta^{\tilde{L}}_{21}
-1.12 \Delta^{\tilde{L}}_{23}\Delta^{\tilde{L}}_{31}
)
\right\}\, ,&
\\
&g_4=6.4\times 10^{-5} \left(\frac{\rm 100GeV}{m_{S}}\right)^2&
\nonumber\\
&
\times \left\{
(1-0.24 \sin 2\beta) \Delta_{21}^{\tilde{L}}
+(-0.81+0.12(\sin2\beta+\cos2\beta))\Delta_{23}^{\tilde{L}}\Delta_{31}^{\tilde{L}}
\right\}\, ,&
\\
&g_6=-1.9 \times 10^{-5} \left(\frac{\rm 100GeV}{m_{S}}\right)^2&
\nonumber\\
&
\times \left\{
(1-0.70 \sin 2\beta) \Delta_{21}^{\tilde{L}}
+(-0.43+0.35(\sin2\beta+\cos2\beta))\Delta_{23}^{\tilde{L}}\Delta_{31}^{\tilde{L}}
\right\}\,,&
\end{eqnarray}
where
\begin{eqnarray}
\Delta^{\tilde{L}}_{ij} \equiv \left(\frac{(\delta m_{\tilde
L}^2)_{ij}}{m_S^2}\right),
\nonumber\\
\Delta^{A_e}_{ij} \equiv \left(\frac{(\delta A_e)_{ij}/Y_{e_i}}
{m_S}\right).
\nonumber
\end{eqnarray}
We remind that $\Delta_{21}^{\tilde{L}}=(\Delta_{12}^{\tilde{L}})^\star.$
The $\sin2\beta$ and $\cos2\beta$ dependences of $g_4$ and $g_6$ 
above are due to $Z$-penguin diagrams. In the branching ratio for
$\mu
\rightarrow 3e$, the contribution from $A_R$ tends to dominate
due to the phase-space integral. Then, assuming that $A_R$ dominates
in ${Br}(\mu^{+} \rightarrow e^{+}e^{+}e^{-})$, the T-odd asymmetry $A_T$
is given by
\begin{eqnarray}
A_T=\frac{{\rm Im}\Big[\Delta_{12}^{\tilde{L}} \Delta_{23}^{\tilde{L}} 
\Delta_{31}^{\tilde{L}} \Big]}{|\Delta_{12}^{\tilde{L}}|^2}
\frac{0.039+0.196 \tan\beta + 0.017/\tan\beta }
     {
\left|(1+2.4\tan\beta)
-
\frac{\Delta_{23}^{\tilde{L}} \Delta_{31}^{\tilde{L}}}{\Delta_{21}^{\tilde{L}}}
(0.64 +1.12\tan\beta)
\right|^2
}\,,
\label{approxAT}
\end{eqnarray}
where we have expanded $\sin2\beta$ and $\cos2\beta$ in terms of 
$\tan\beta$.  Also, in writing
(\ref{approxAT}), we have taken
${\Delta}_{21}^{A_e}={\Delta}_{23}^{A_e}=0$, for simplicity.

We see explicitly how $A_T$ (\ref{approxAT}) depends on the Jarlskog
invariant $J_{\tilde L}$ (\ref{Jl}), and it is apparent how analogous
invariants $J_{A_{12}}$, etc., with one or more
${\Delta}^{\tilde{L}}_{ij} \to {\Delta}^{A_e}_{ij}$ could also
contribute.  We see that $A_T$ could in principle reach $\sim 10\%$.
However, if ${\rm Im}[\Delta_{12}^{\tilde{L}} \Delta_{23}^{\tilde{L}}
\Delta_{31}^{\tilde{L}}] \ll |\Delta_{12}^{\tilde{L}}|^2$, as one
might expect, or if $\tan\beta\gg 1$, $A_T$ is suppressed.
However, we stress that \Eq{approxAT} is approximately correct 
only for \rfn{limit1} and cannot be used to predict $A_T$ in
more general cases which will be considered in the next Section.

Next, we present approximate formulae for the effective coupling in
(\ref{edm_eff}) in the specific case (\ref{limit1}). Since relative phases
between $(m^2_{\tilde{L}})$, $(m^2_{\tilde{E}})$ and $(A_{e})$ contribute
to the electric dipole moments, non-vanishing contributions to $d_l$ come
only from slepton diagrams, not sneutrino diagrams. From the explicit
formula (\ref{dlc}), it is also clear that the sneutrino diagrams do not
give a non-vanishing value, since they depend on
$(U_\nu)_{Xi}^\star(U_\nu)_{Xi}$, and the CP-violating phases exist only
in the mixing matrices of the sleptons, in our approximation.

Since it depends on the neutrino model which contribution is
dominant in the electric dipole moments, as shown in the previous Section,
we first show the general formula for $d_l^{\tilde{\chi^0}}$
in the limit (\ref{limit1}):
\begin{eqnarray}
\frac{d_l^{\tilde{\chi^0}}}{e}
&=&
\frac{g_Y^2}{(4 \pi)^2}
\frac{1}{m_S}
\mrm{Im}\left[
\sum_{N=1} \sum_{i_1,\cdots,i_N} 
c_N \Delta_{l i_1}\Delta_{i_1 i_2}\cdots \Delta_{i_N l}
\right]
.
\end{eqnarray}
Here, $\Delta_{ij}$ is the flavor-dependent part of the slepton mass
matrix, which normalized by $m_S^2$, and includes parts generated by
the renormalization as well as tree-level parts. At least one of the
$\Delta_{ij}$ in each product term involves the left-right mixing of a
slepton. The coefficients $c_i$ are:
\begin{eqnarray}
c_1=-\frac1{12},~
c_2= \frac1{20},~
c_3=-\frac1{30},~
c_4= \frac1{42},~
c_5=-\frac1{56}.
\end{eqnarray}
When the heavy singlet-neutrino masses are not almost degenerate, and
corrections to the soft supersymmetry-breaking terms and the relative
phases among $(m^2_{\tilde{L}})$ and $(A_{e})$ are generated at
${\cal O}(\log^2\frac{M_{GUT}}{M_N})$, the approximate formula becomes
\begin{eqnarray}
\frac{d_l^{\tilde{\chi^0}}}{e}
&=&
\frac{g_Y^2}{(4 \pi)^2} \frac{m_l}{m_S^2}
\left(
\frac{1}{42} \tan\beta^2
+\frac{1}{21} \tan\beta
-\frac{1}{105}
\right) 
\sum_{j} \mrm{Im}\left[\Delta^{A_e}_{lj}
\frac{m_{l_j}^2}{m_S^2}
\Delta^{\tilde{L}}_{jl}
\right]
\nonumber\\
&&
-\frac1{30} \frac{g_Y^2}{(4 \pi)^2} \frac{m_l}{m_S^2}
\sum_{j,k} \mrm{Im}\left[
\Delta^{A_e}_{lj}
\frac{m_{l_j}^2}{m_S^2}
(\Delta^{A_e}_{kj})^\star
\Delta^{A_e}_{kl}
\right]
,
\end{eqnarray}
where the first term is proportional to (\ref{jedm2}) and the second
to (\ref{jedm2p}).

When the heavy singlet-neutrino masses are almost degenerate and the
correction to $\mrm{Im}[A_e]_{ii}$, proportional to $\log
\frac{M_{N_i}}{M_{N_j}} \log\frac{M_{GUT}}{M_{N}}$, is dominant in the
electric dipole moments, the approximate formula is
\begin{eqnarray}
\frac{d_l^{\tilde{\chi^0}}}{e}
&=&
-\frac{1}{12} \frac{g_Y^2}{(4 \pi)^2} \frac{m_l}{m_S^2}
\mrm{Im}\left[
\Delta^{A_e}_{ll}
\right] .
\end{eqnarray}

\section{Numerical Analysis}

\subsection{Calculational Procedure}

We first fix the gauge couplings,
charged-lepton and quark Yukawa couplings and $\tb$ (which is a free
parameter) at the scale $M_Z$, and then run them with the two-loop
MSSM renormalization-group equations up to the scale $M_N.$ At $M_N$, we
introduce the heavy singlet neutrinos, fixing their masses, the
light-neutrino masses and mixings according to the oscillation data. We
then choose the matrix $R$ and calculate ${ Y_\nu}$ according to
(\ref{Ynu}).
Subsequently, we
run all the Yukawa-coupling matrices from $M_N$ to $M_{GUT}$ using the
one-loop renormalization-group equations~\cite{h1}. At $M_{GUT}$ we assume
universal boundary conditions (\ref{mssmbound})  for the soft
supersymmetry-breaking terms. We then run all the soft
supersymmetry-breaking masses and Yukawa matrices back to $M_N$, where the
heavy singlet neutrinos and sneutrinos decouple. The soft
supersymmetry-breaking mass matrices at low energies are obtained finally
by running all the MSSM parameters back down to $M_Z.$ We use the
electroweak symmetry-breaking conditions to fix the magnitude of the Higgs 
mixing parameter $\mu$, taking its sign positive as motivated by
$g_\mu-2.$ This sign is also consistent with the bounds from $b\to
s\gamma.$ Then we calculate the squark, slepton, chargino and
neutralino mass matrices and finally LFV rates, the T-odd asymmetry 
$A_T$ in polarized
$\mu^{+} \rightarrow e^{+}e^{+}e^{-}$ decays and the EDMs of the electron 
and muon, 
for chosen values of the input parameters. 

\subsection{Illustrative Results}

In our numerical examples, we take~\cite{3nufit} $\Delta
m^2_{atm}=3.5\times 10^{-3}$ eV$^2$ with $\sin\theta_{23}=0.7$ for
atmospheric neutrinos, and $\Delta m^2_{sol}=5.0\times 10^{-5}$ eV$^2$
with $\tan^2\theta_{12}=0.45$ for solar neutrinos, corresponding to the
LMA solution. Also, we fix $\tan^2\theta_{13}=0.055$, which
is the largest value allowed by the global three-neutrino data fit.
We parametrize the orthogonal matrix $R$ in the form
\bea
\label{R}
R=\pmatrix{\hat c_2\hat c_3 & -\hat c_1\hat s_3-\hat s_1\hat s_2\hat
c_3  & \hat s_1\hat s_3-\hat c_1\hat s_2\hat c_3 \cr  \hat c_2\hat s_3
& \hat c_1\hat c_3-\hat s_1\hat s_2\hat s_3   & -\hat s_1\hat c_3-\hat
c_1\hat s_2\hat s_3 \cr \hat s_2  & \hat s_1\hat c_2 & \hat c_1\hat
c_2\cr}\;,  
\eea
where $\hat \theta_1$, $\hat \theta_2$, $\hat \theta_3$ are arbitrary
complex angles. Using a generic $R$ as input is crucial
for neutrino phenomenology. We stress the following: 
\begin{enumerate}
\item[{\it (i)}] 
Because $R$ is a complex orthogonal matrix, the values of its entries are
not restricted to any small range, but rather are {\it exponential}
functions of complex numbers. This implies via (\ref{Ynu}) that, for a
suitable choice of $\hat \theta_{1,2,3}$, all the elements of ${
Y}_\nu$ can be large even if one starts with small (for example
hierarchical) neutrino masses;
\item[{\it (ii)}] 
Large imaginary components are in general present in every entry of ${
Y}_\nu$. Since $Y^\dagger_\nu Y_\nu$ depends on the phases of both $R$ and
$U$, as seen in (\ref{y+y2}), sizeable CP-violating
effects may be induced. However, if $R=1$, only $\delta$ in $U$
contributes to $Y^\dagger_\nu Y_\nu$.
\end{enumerate} 

We first calculate $d_e$ and $d_\mu$ using (\ref{dl+}) and
(\ref{dl0}), and study their possible ranges in our model by scanning
over the allowed values of the free parameters.  As expected, $d_e$ and
$d_\mu$ are very small. For degenerate right-handed neutrinos,
typically $d_\mu$ does not exceed $10^{-31}$ e~cm for the values of
the parameters consistent with the bounds on $\mu\to e\gamma,$ and is
therefore many orders of magnitude below the sensitivity of any
planned experiment. The approximate relation between muon and electron
electric dipole moments in our model, $d_\mu/d_e\approx -m_\mu/m_e,$
holds very well in this case. However, in the case of non-degenerate
right-handed neutrinos, the $\log (M_{N_i}/M_{N_j})$ effects introduce
a dependence of the electric dipole moments on the leptogenesis
phases, as discussed previously. These new contributions may change the
size of electric dipole moments by  a few orders of magnitude. 
In particular, similarly to other supersymmetric
models~\cite{strumia}, the sign of $d_\mu/d_e$ may be
altered and the naive relation $d_\mu/d_e\approx -m_\mu/m_e$ may be
violated by a large factor.
Detailed analyses of the electric dipole moments in the case of
non-degenerate heavy neutrinos will be presented elsewhere \cite{us2}.

On the other hand, the situation with the T-odd asymmetry in polarized
$\mu^{+} \rightarrow e^{+}e^{+}e^{-}$ decays can be different. As
explained in the previous Section, the decay $\mu^{+} \rightarrow
e^{+}e^{+}e^{-}$ receives contributions from box diagrams and photonic
penguin diagrams, with the latter usually dominating. However, if there
are cancellations in the dipole-moment-type $\mu-e-\gamma$ vertex, the box
and penguin contributions to $\mu^{+} \rightarrow e^{+}e^{+}e^{-}$ may
become comparable. In that case, if there are large CP-violating phases
present in the slepton mass matrices, the T-odd asymmetry $A_T$ can be
large.  This implies an anti-correlation between ${Br}(\mu\to e\gamma)$ and
$A_T:$ the latter can be large only if the former is suppressed.

To illustrate how such a cancellation might come about, consider the
two-generation example described at the end of Section 2, which also
applies in the full three-generation case if $m_{\nu_3} \ll
m_{\nu_{1,2}}$:  we assume for simplicity that also $m_{\nu_2} \ll
m_{\nu_{1}}$.  We see from (\ref{Rey+y}) that $\mbox{Re}\Big[\left( {
Y_\nu}^\dagger { Y_\nu} \right)^{12} \Big]$ is suppressed if
\begin{equation}
(M_1 + M_2) \cos 2 \theta_r \approx (M_1 - M_2) {\rm cosh} 2 \theta_i,
\label{cancelRe}
\end{equation}
and that the smaller quantity $\mbox{Im}\Big[\left( { Y_\nu}^\dagger {
Y_\nu} \right)^{12} \Big]$ is also suppressed if
\begin{equation}
(M_1+M_2) \cos\phi {\rm sinh} 2 \theta_i \approx
(M_1-M_2) \sin\phi \sin 2 \theta_r.
\label{cancelIm}
\end{equation}
Both the conditions (\ref{cancelRe},\ref{cancelIm}) may in principle be
satisfied simultaneously for suitably tuned values of $\theta_i,
\theta_r$ and $\phi$.

In a general and more physical
case, the cancellation in ${Br}(\mu\to e\gamma)$ is much more 
complicated and depends on all free parameters, namely
the phases and mixing angles in
the matrices $U$ and $R,$ the values of the light- and heavy- neutrino 
masses,
the choice of the soft supersymmetry-breaking initial conditions,
details of the renormalization-group
running procedure, etc.. However, the fact that such cancellations
occur is robust and qualitatively well understood.
To give a representative numerical example, we choose
hierarchical neutrino masses with $m_{\nu_1}=0.028$ eV,
$M_{N_1}=1.2\times 10^{15}$ GeV,
$M_{N_2}=1.5\times 10^{15}$ GeV,
$M_{N_3}=3\times 10^{14}$ GeV,
$\delta=\pi/2,$ $\phi_1=-1,$ 
$\hat \theta_1=0.3 i,$
$\hat \theta_2=0.5 i,$ 
$\hat \theta_3=0.1 i,$
and use the same neutrino oscillation parameters as above.
We run the renormalization-group equations with right-handed 
neutrinos down to the scale $M_N=3\times 10^{14}$ GeV.

\begin{figure}[t]
\centerline{\epsfxsize = 0.5\textwidth \epsffile{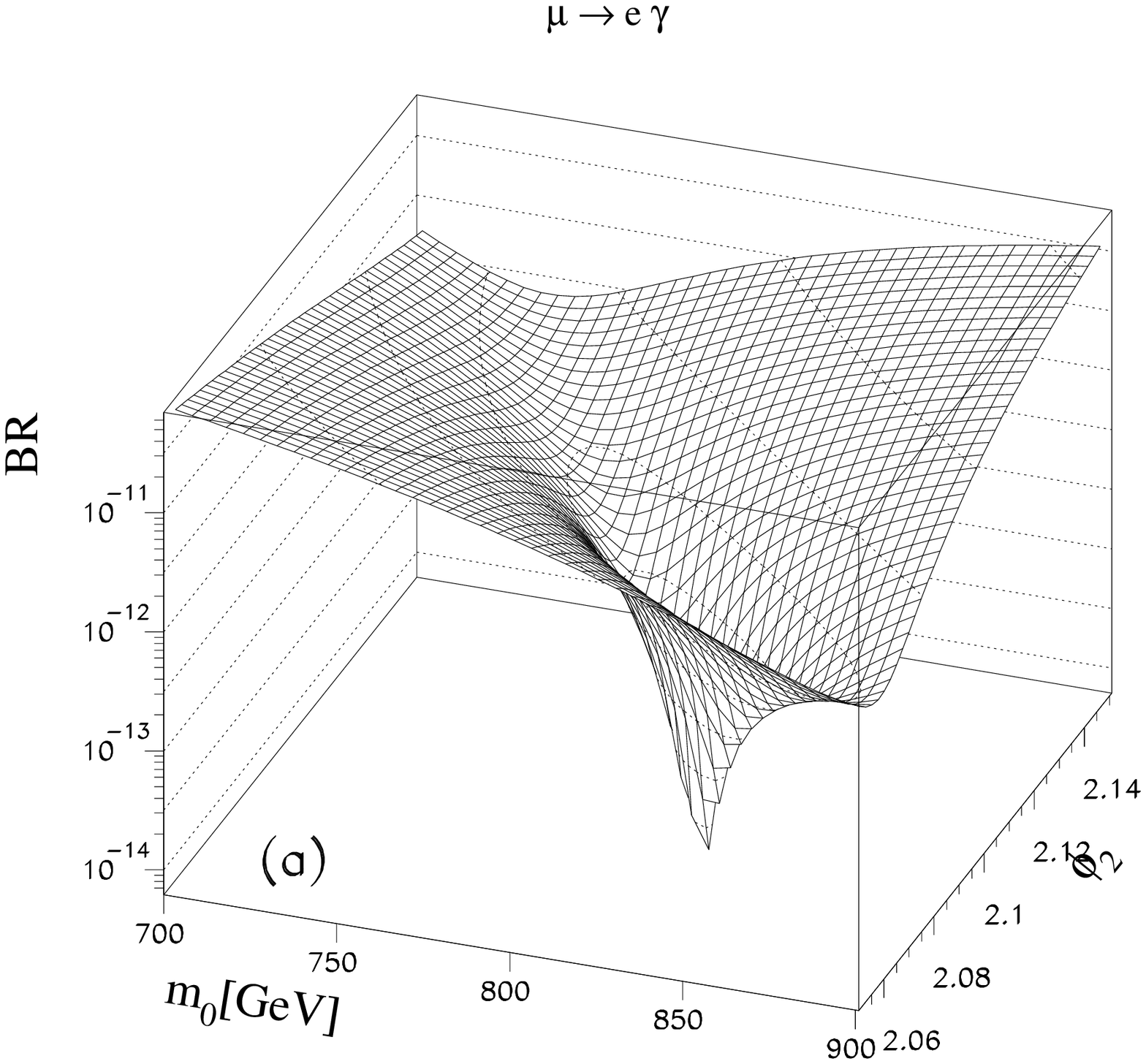} 
\hfill \epsfxsize = 0.5\textwidth \epsffile{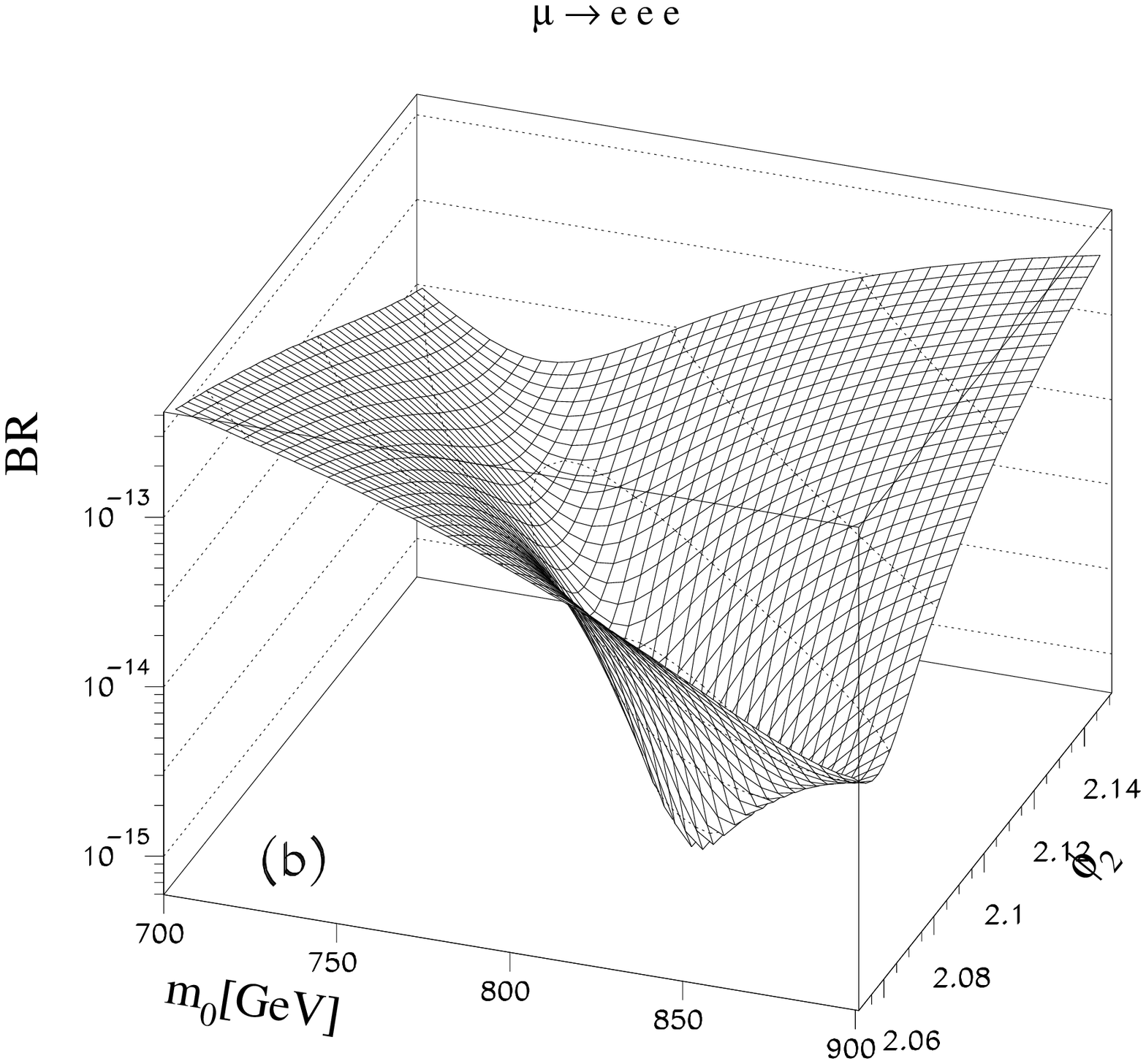} }
\caption{\it Branching ratios for the decays
(a) $\mu^{+} \rightarrow e^{+}\gamma$   and 
(b) $\mu^{+} \rightarrow e^{+}e^{+}e^{-}$   
as functions of the common soft mass $m_0$ and the Majorana phase $\phi_2$, 
for the fixed choice of neutrino parameters described in the text.
\vspace*{0.5cm}}
\label{fig1}
\end{figure}

\begin{figure}[htb]
\centerline{\epsfxsize = 0.6\textwidth \epsffile{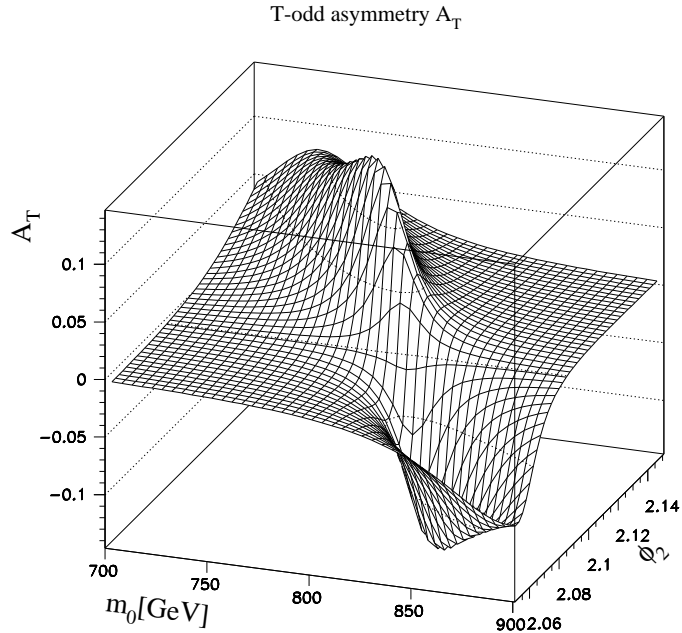} 
}
\caption{\it The T-odd asymmetry $A_T$ in polarized 
$\mu^{+} \rightarrow e^{+}e^{+}e^{-}$
decay for the same set of parameters as in Fig.~\ref{fig1}.
\vspace*{0.5cm}}
\label{fig2}
\end{figure}

Fixing these parameters, we first choose $m_{1/2}=500$ GeV, 
$A_0=0$ GeV and $\tb=20$ and scan over the remaining free parameters $m_0$ and 
$\phi_2.$ In Fig. \ref{fig1} we plot the branching
ratios of the decays $\mu^{+} \rightarrow e^{+}\gamma$ and $\mu^{+}
\rightarrow e^{+}e^{+}e^{-}$ as functions of the common sfermion soft mass
parameter $m_0$ and the Majorana phase $\phi_2.$ 
For a large region in the plotted parameter space, 
${Br}(\mu^{+} \to e^{+}\gamma)$ is below the present experimental bound,
whilst ${Br}(\mu^{+} \to e^{+}e^{+}e^{-})$ does not at present
impose any constraints.
The branching ratios in Fig. \ref{fig1} are correlated, implying that the
photonic penguin diagrams are dominating also in $\mu^{+} \rightarrow
e^{+}e^{+}e^{-}.$ For the plotted values, both rare decays should be observed
in the planned experiments.

The T-odd asymmetry $A_T$ is plotted in Fig. \ref{fig2} for the same
set of parameters as in Fig. \ref{fig1}. Comparison with Fig. \ref{fig1}
shows that $A_T$ is strongly anti-correlated with the branching ratios.  
For the region with the deepest cancellation, the T-odd asymmetry is negative.
Whilst its absolute value may exceed 10\%, the branching ratio of 
$\mu^{+} \to e^{+}e^{+}e^{-}$ is relatively small in that region, making
its precise determination difficult. However, for  large positive
$A_T$, ${Br}(\mu^{+} \to e^{+}e^{+}e^{-})$ exceeds the $10^{-14}$ level,
implying that several hundred events could be observed in the planned 
experiments.
Since the experimental sensitivity to $A_T$ should scale as
$1/\sqrt{N_{events}}$, these future experiments
would be able to measure a non-zero value of $A_T$ for large ranges of 
$m_0$ and $\phi_2$ in Fig. \ref{fig2}.

\begin{figure}[t]
\centerline{\epsfxsize = 0.5\textwidth \epsffile{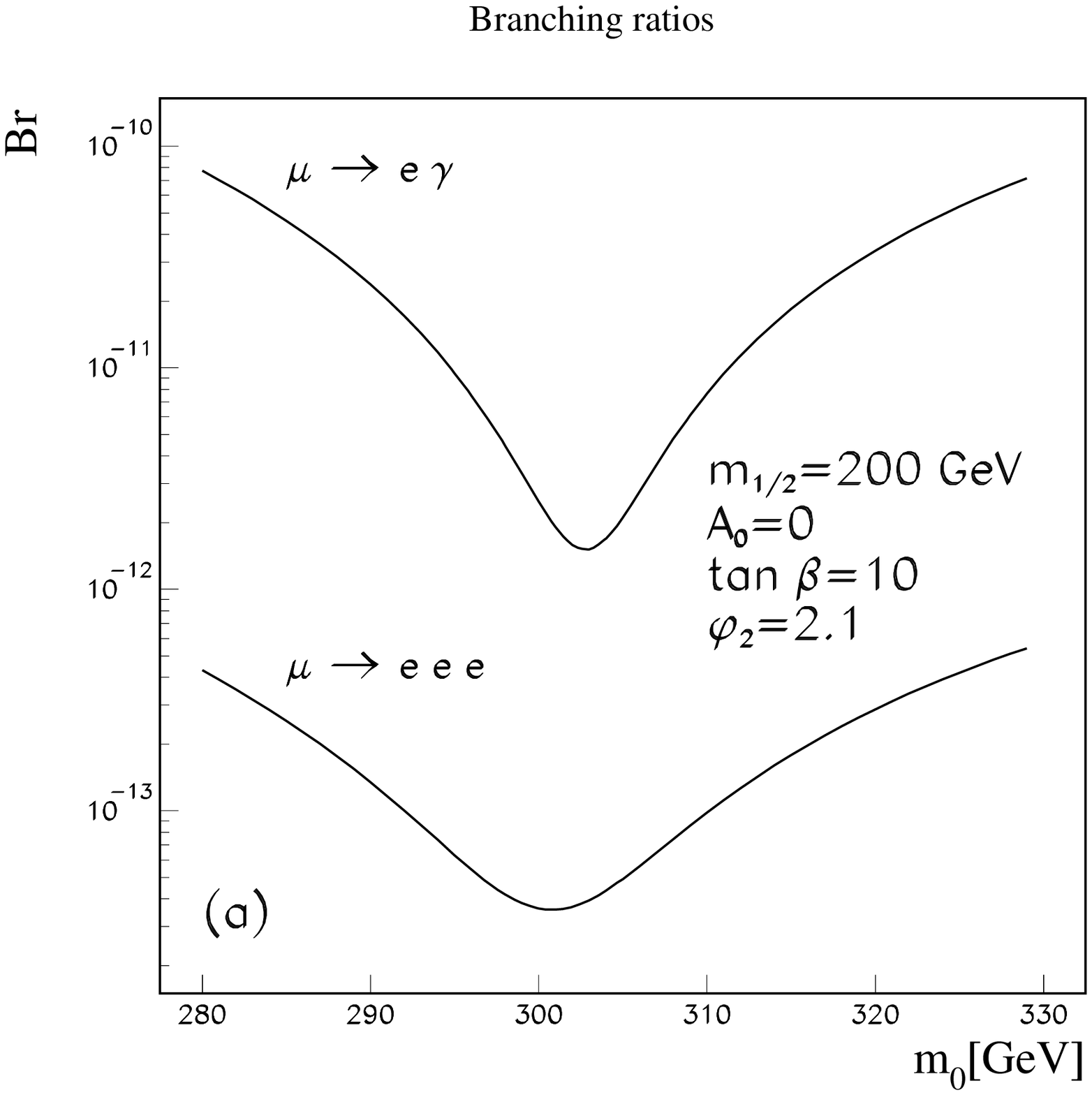} 
\hfill \epsfxsize = 0.5\textwidth \epsffile{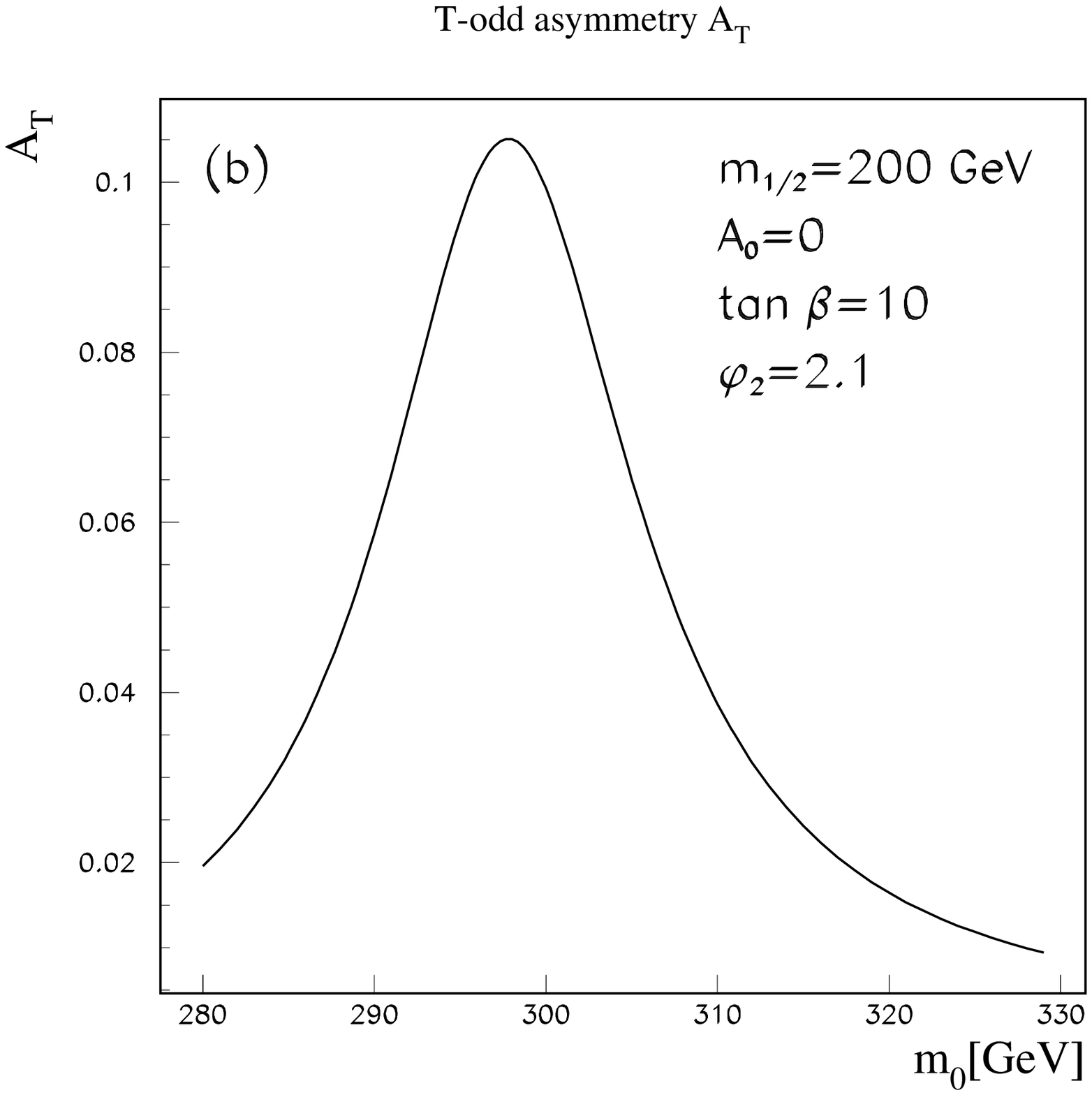} }
\caption{\it (a) Branching ratios for the decays
$\mu^{+} \rightarrow e^{+}\gamma$ and $\mu^{+} \rightarrow e^{+}e^{+}e^{-}$
and (b) the T-odd asymmetry $A_T$ in $\mu^{+} \rightarrow e^{+}e^{+}e^{-}$
decay, as functions of the common soft mass $m_0$, for the fixed choice of
neutrino parameters described in the text.
\vspace*{0.5cm}}
\label{fig3}
\end{figure}
\begin{figure}[t]
\centerline{
\epsfxsize = 0.5\textwidth \epsffile{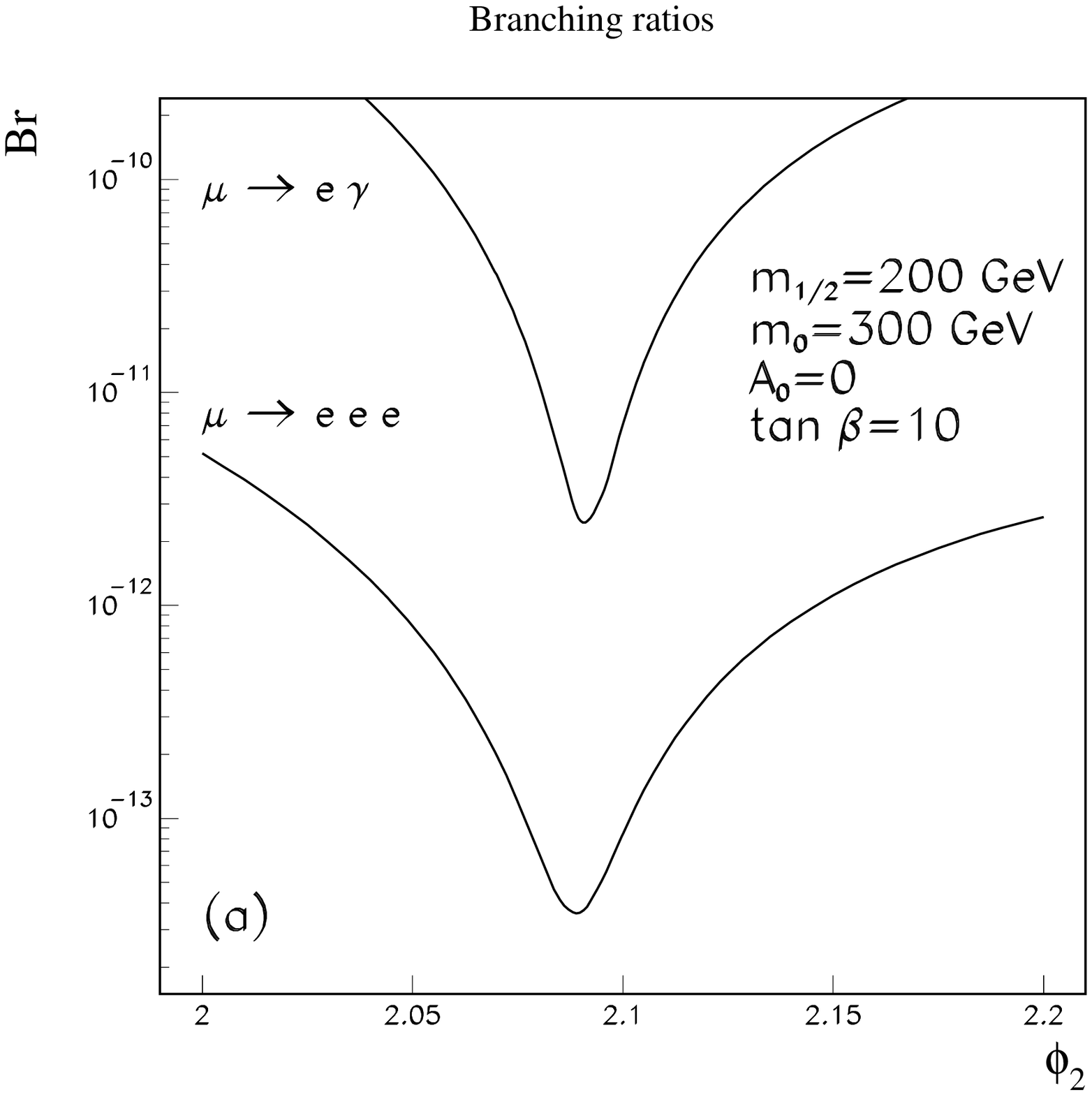} 
\hfill
\epsfxsize = 0.5\textwidth \epsffile{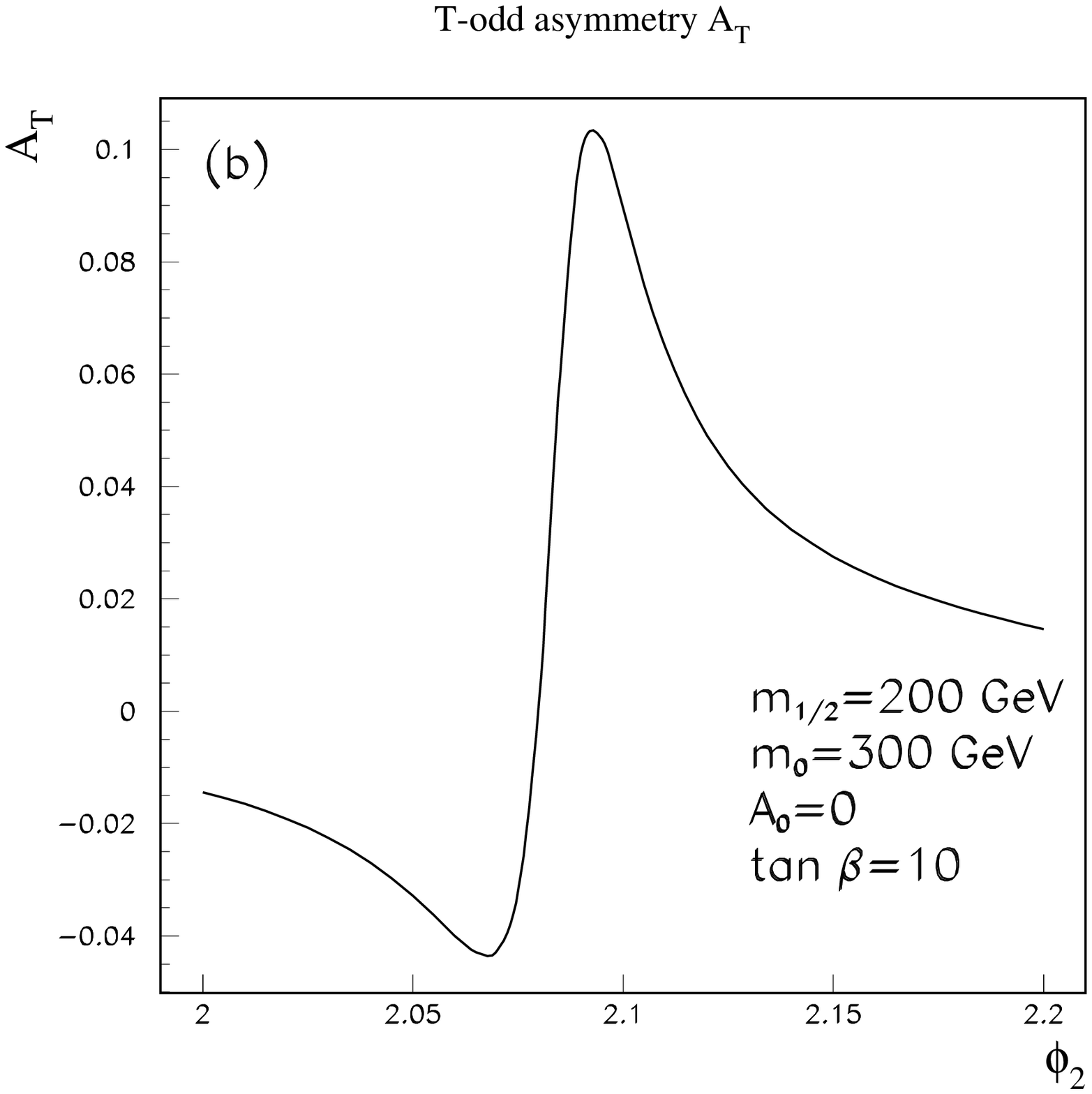}
}
\caption{\it 
(a) Branching ratios of the decays 
$\mu^{+} \rightarrow e^{+}\gamma$ and $\mu^{+} \rightarrow e^{+}e^{+}e^{-}$
and (b) the T-odd asymmetry $A_T$ in $\mu^{+} \rightarrow e^{+}e^{+}e^{-}$
decay, as functions of the Majorana phase $\phi_2$ for  $m_0=300$ GeV.
All other parameters are fixed as in Fig. \ref{fig3}.
\vspace*{0.5cm}}
\label{fig4}
\end{figure}

Whilst the long-baseline oscillation experiments at neutrino factories
will  measure  the phase $\delta$ in (\ref{V}), we stress here again that
the T-odd asymmetry $A_T$  depends on all the phases in the matrices
$U$ and $R$. Thus different combinations of phases in the Yukawa
matrix ${ Y_\nu}$ are probed in the neutrino oscillation and
stopped muon experiments. This point is made explicitly by the dependence
of $A_T$ on the Majorana phase $\phi_2$ in Fig. \ref{fig2}.

We have chosen the neutrino parameters in such a way that the absolute values
of all the neutrino Yukawa couplings are close to unity. This induces large
rates of LFV and CP violation. However, due to the cancellations in the
photonic penguin diagrams, the induced ${Br}(\mu^{+} \to e^{+}\gamma)$
can be consistent with the current experimental bounds even for
relatively small sparticle masses. In Fig. \ref{fig3} we plot 
(a) ${Br}(\mu^{+} \to e^{+}\gamma)$ and ${Br}(\mu^{+} \to e^{+}e^{+}e^{-})$ and
(b) $A_T$ as functions of $m_0$ for fixed 
$m_{1/2}=200$ GeV, $A_0=0$ GeV, $\tb=10$ and $\phi_2=2.1$.
For a small region  around  $m_0=300$ GeV, ${Br}(\mu^{+} \to e^{+}\gamma)$
is below the present limit. At the same time, the T-odd asymmetry in 
$\mu^{+} \to e^{+}e^{+}e^{-}$ may be as large as 10\% for the
allowed values of $m_0.$ The dependence of the branching ratios
and the T-odd asymmetry on the Majorana phase $\phi_2$ is demonstrated
in    Fig. \ref{fig4}. Again, large $A_T$ is expected if the 
decay $\mu^{+} \to e^{+}\gamma$ is suppressed due to the cancellation.

\begin{figure}[t]
\centerline{
\epsfxsize = 0.5\textwidth \epsffile{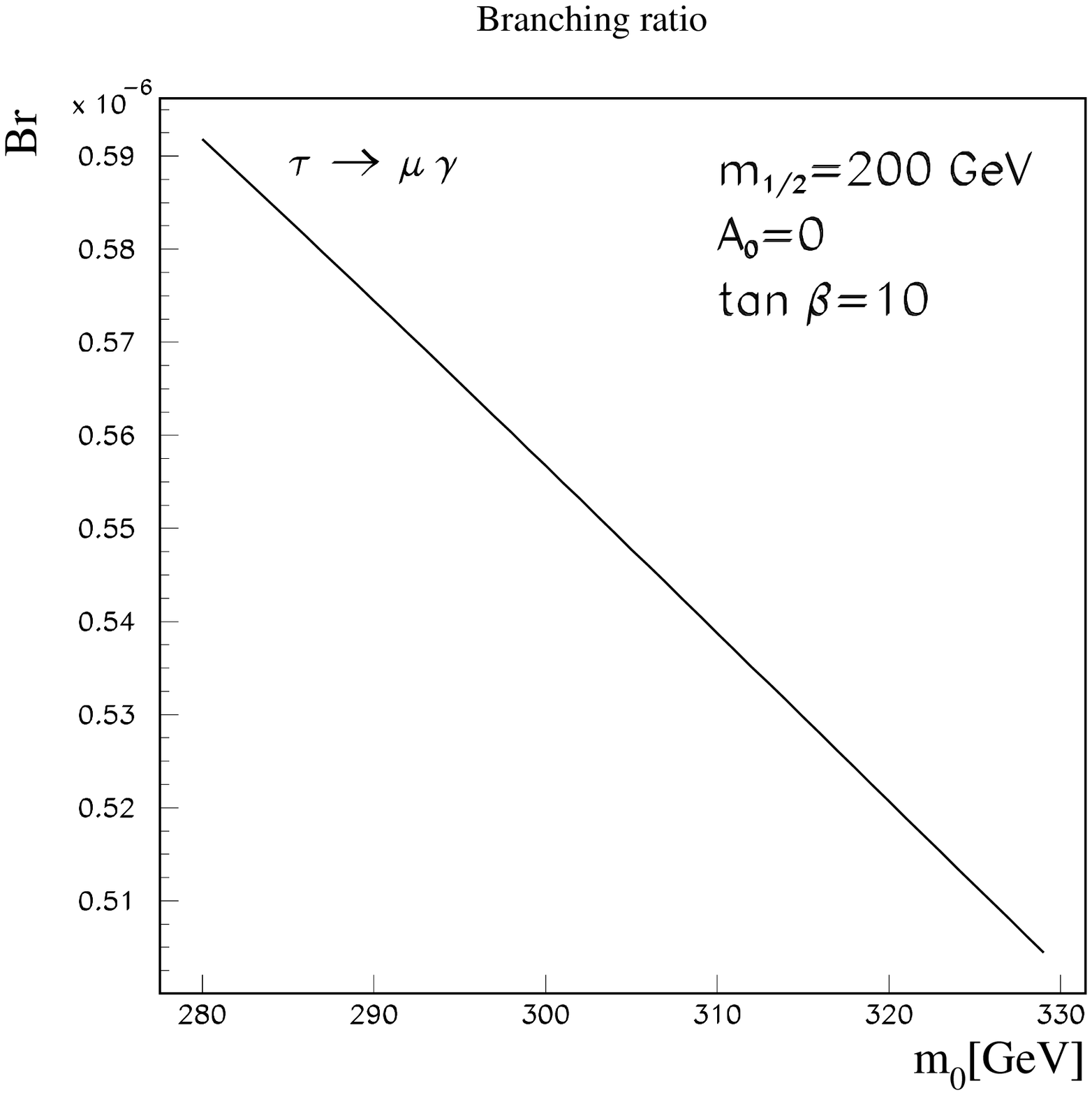} 
}
\caption{\it 
Branching ratio of the decay
$\tau^{+} \rightarrow \mu^{+}\gamma$ as a function of $m_0.$
All other parameters are fixed as in Fig. \ref{fig3}.
\vspace*{0.5cm}}
\label{fig5}
\end{figure}

Finally, we note that the branching ratio of the decay 
$\tau^{+} \to \mu^{+}\gamma$ {\it does not} have cancellations in the 
parameter region considered, as seen in Fig. \ref{fig5}.
Therefore, in this scenario ${Br}(\tau^{+} \to \mu^{+}\gamma)$
might be just below the present experimental bound and discoverable at 
the LHC or the B factories. As the decays $\tau \to \ell \ell \ell$ are 
suppressed relative to
$\tau^{+} \to \mu^{+}\gamma,$ detailed quantitative studies of them are
beyond the interest of the present work.

\section{Discussion and Conclusions}

We have seen in this paper that the T-odd asymmetry $A_T$ in polarized
$\mu^{+} \rightarrow e^{+}e^{+}e^{-}$ decay may offer the best prospects
for studying in the laboratory CP-violating effects in the minimal
supersymmetric seesaw model. On the other hand, the electric dipole
moments of the electron and muon are suppressed in the minimal
supersymmetric seesaw scenario discussed here. This is because the
CP-violating phases are induced by renormalization-group running only in
the off-diagonal entries in the slepton mass matrices, as already
discussed in~\cite{acfh}. The naive relation $d_\mu/d_e\approx -m_\mu/m_e$ 
holds very well in the case of degenerate right-handed neutrinos.
In the case of non-degenerate right-handed neutrinos, logarithmic effects
arising from $\log (M_{N_i}/M_{N_j})$ introduce a 
dependence on the leptogenesis
phases. These new contributions may become dominant and 
the naive relation  $d_\mu/d_e\approx -m_\mu/m_e$ is badly violated.
In the most optimistic case, the electric dipole moments  may
approach the level observable at the proposed experiments \cite{us2}.

The possibility of measuring a non-zero $A_T$ could have far-reaching
consequences, since it provides complementary information on the
CP-violating phases in the neutrino Yukawa matrix $Y_\nu$.  As has been
discussed, $A_T$ depends in the leading-logarithmic approximation on a
single combination of light-neutrino phases and the three phases in 
$Y_\nu^\dagger Y_\nu$ that
contribute to leptogenesis, whereas the CP-violating phases in the
light-neutrino effective mass matrix depend on the phases in $Y_\nu
Y_\nu^\dagger$, that do not contribute to leptogenesis.

This is one reason why $A_T$ may be observable even if CP violation is
undetectable in neutrino oscillations. We recall also that the latter is
in practice observable only if the neutrino masses and mixing angles are
favourable.  For example, if $U_{e3}\approx 0$ and/or $\Delta m^2_{sol}$
is small, as in the case of vacuum oscillations, CP violation is
unobservable using neutrino factories. However, as seen in (\ref{Ynu}),
(some of) the Yukawa couplings in $Y_\nu$ may still be large and
imaginary, implying that $A_T$ might be large.

On the other hand, a large value of $A_T$ requires cancellations in the
slepton-induced $\mu-e-\gamma^\star$ vertex, which happens only in a
restricted region of the parameter space. The asymmetry $A_T$ is
anti-correlated with the branching ratio of $\mu\rightarrow e \gamma$, and
it can reach $\sim 10 \%$ if $\mu\rightarrow e \gamma$ is suppressed. The
asymmetry $A_T$ may be measurable in planned high-intensity stopped-muon
experiments, which aim at a sensitivity to ${Br}(\mu\rightarrow eee) \sim
10^{-16}.$

In the case of $\tau\rightarrow \mu \gamma$ the cancellation does not happen 
for the same parameters as in $\mu\rightarrow e \gamma.$ Therefore,
in the scenario considered in this paper ${Br}(\tau\rightarrow \mu \gamma)$
is large and can be observed at the LHC experiments.

It is interesting to review what we would learn if non-vanishing $A_T$ 
were observed. The T-odd asymmetry $A_T$ is approximately proportional to the
CP invariant (\ref{Jl}). We recall that, if $Y_\nu$ has a hierarchical 
structure, the
off-diagonal components of the left-handed slepton mass matrix are given
by
\bea
\left(m_{\tilde L}^2\right)_{12} & \propto & 
  (Y_{\nu_3}^D)^2 X_{31} X_{32}^\star 
  +(Y_{\nu_2}^D)^2 X_{21} X_{22}^\star \, , 
\nn \\
\left(m_{\tilde L}^2\right)_{23} \left(m_{\tilde L}^2\right)_{31} 
& \propto& 
 (Y_{\nu_3}^D)^4 |X_{33}|^2
X_{32} X_{31}^\star\, .
\nn 
\eea
The Jarlskog invariant $J_{\tilde{L}}$ may have a sizeable value if 
$(Y^D_{\nu_3})^2 X_{31}
X_{32}^\star$ and $(Y^D_{\nu_2})^2 X_{21} X_{22}^\star$ are
comparable. Thus, if non-vanishing $A_T$ is observed, the generation
structure in the neutrino Yukawa coupling may
be constrained, as well as the CP-violating phase.

In conclusion: searching for CP violation in lepton-flavour-violating
processes is a possibility that should not be neglected, since it provides
information complementary to that provided by neutrino oscillation
experiments. In particular, $A_T$ may be measurable even if CP violation
is unobservable in neutrino oscillations.

\vskip 0.5in
\vbox{
\noindent{ {\bf Acknowledgments} } \\
\noindent  
We thank L. N.~Chang, A. De~Gouvea, B.~Gavela, C.~Gonzalez-Garcia, 
B.~Kayser, S.~Khalil, N.~Sakai, A.~Strumia and F. Vissani for 
enlightening discussions.  This work is partially supported by EU TMR
contract No.  HPMF-CT-2000-00460, ESF grant No. 3832, and the
Grant-in-Aid for Scientific Research from the Ministry of Education,
Science, Sports and Culture of Japan, on Priority Area 707
`Supersymmetry and Unified Theory of Elementary Particles'.

}

\end{document}